\documentclass{jpp}
\usepackage{graphicx}
\usepackage{subcaption}
\usepackage{wrapfig}
\usepackage{tikz}
\usepackage{pgfplots}
\pgfplotsset{compat=1.18}
\usetikzlibrary{angles, quotes, arrows.meta, positioning, decorations.pathmorphing, decorations.markings, calc, 3d, shapes.arrows}
\usepackage{xcolor}
\usepackage[utf8]{inputenc}
\usepackage[T1]{fontenc}
\usepackage{dirtytalk}
\usepackage{scalerel,stackengine}

\usepackage{amsmath}
\usepackage{derivative}
\usepackage{bm}
\usepackage{physics}
\AtBeginDocument{\RenewCommandCopy\qty\SI}
\usepackage{tabularx}
\usepackage{comment}
\usepackage{booktabs}
\usepackage{siunitx}

\usepackage[dvipsnames]{xcolor} 

\usepackage[hidelinks]{hyperref} 
\usepackage{bookmark}

\usepackage{cleveref}
\crefformat{section}{\S#2#1#3} 
\crefformat{subsection}{\S#2#1#3}
\crefformat{subsubsection}{\S#2#1#3}

\hypersetup{
  colorlinks=true,
  linkcolor=blue,
  citecolor=blue,
  urlcolor=blue,
  bookmarksopen=true
}
 
\usepackage[table]{xcolor} 
\definecolor{improved}{RGB}{198, 239, 206}
\definecolor{same}{RGB}{242, 242, 242}
\definecolor{worse}{RGB}{255, 199, 206}

\shorttitle{Parker spiral and the reflection-driven turbulence}
\shortauthor{K. Abbas and J. Squire}

\title{The influence of Parker spiral on the reflection-driven turbulence}

\author{Khurram Abbas\aff{1}
  \corresp{\email{khurram.abbas@postgrad.otago.ac.nz}},
  Jonathan Squire\aff{1}}

\affiliation{\aff{1}Department of Physics, University of Otago,
730 Cumberland St., Dunedin 9016, New Zealand}

\begin{document}

\maketitle

\begin{abstract}
The solar wind is observed to undergo substantial heating as it expands through the heliosphere, with measured temperature profiles exceeding those expected from adiabatic cooling. A plausible source of this heating is reflection-driven turbulence (RDT), in which gradients in the background Alfvén speed partially reflect outward-propagating Alfvén waves, seeding counter-propagating fluctuations that interact and dissipate via turbulence. Previous RDT models assume a radial background magnetic field, but at larger radii the interplanetary field is known to be twisted into the Parker Spiral (PS). Here, we generalize RDT phenomenology to include a PS, using three-dimensional expanding-box magnetohydrodynamic (MHD) simulations to test the ideas and compare the resulting turbulence to the radial-background-field case. We argue that the underlying RDT dynamics remain broadly similar with a PS, but the controlling scales change: as the azimuthal field grows it \say{cuts across} perpendicularly stretched, pancake-like eddies, producing outer scales perpendicular to the magnetic field that are much smaller than in the radial-background case. Consequently, the outer-scale nonlinear turnover time increases more slowly with heliocentric distance in PS geometry, weakening the tendency (seen in radial-background models) for the cascade to ‘freeze’ into quasi-static, magnetically dominated structures. This allows the system to dissipate a larger fraction of the fluctuation energy as heat, also implying that the turbulence remains strongly imbalanced (with high normalized cross-helicity) out to larger heliocentric distances. We complement our heating results with a detailed characterization of the turbulence (e.g., spectra, switchbacks, and compressive fractions) providing a set of concrete predictions for comparison with spacecraft observations.
\end{abstract}

\section{Introduction}
The solar corona and the solar wind exhibit temperature and velocity profiles that cannot be fully explained by \citet{Parker1958}’s simple isothermal hydrodynamic model.
The mechanisms that supply the  energy required to explain the differences remain debated \citep{DeMoortel2015, Cranmer2017, Chandran2021}. A prominent idea is that motions in the photosphere launch Alfvénic waves into open flux tubes. These waves travel outward and somehow dissipate, heating the plasma \citep{Cranmer2009, DePontieu2007}, both raising plasma temperature and increasing magnetic or wave pressure, thereby contributing to wind acceleration \citep{Tu1987,Cranmer2007}. Observational support comes from measurements of ubiquitous Alfvénic fluctuations in fast solar wind streams, as reported by early spacecraft \citep{Belcher1971, Tu1995} and consolidated in later reviews \citep{Bruno2013}. Recent in-situ data indicate that large-amplitude Alfvén waves can carry substantial energy for coronal heating and wind acceleration \citep{Halekas2023,Rivera2024}.

However, a fundamental theoretical limitation arises because in a homogeneous medium, Alfvén waves propagating in a single direction do not interact nonlinearly and thus cannot by themselves sustain a turbulent cascade that dissipate their energy \citep{Kraichnan1965,Barnes1974}. 
In the solar wind, counter-propagating fluctuations can be generated by partial reflection of outward waves from background Alfvén-speed gradients, a process termed \say{reflection-driven turbulence} (RDT) \citep{Velli1989,Velli1993,Matthaeus1999, Verdini2007}. Numerical and analytical models suggest this mechanism can supply a significant fraction of the heating necessary to maintain fast solar wind streams \citep{Verdini2009,Chandran2009, Chandran2019, Chandran2021}.
We note that when compressibility is included, parametric decay instability can also generate counter-propagating and compressible fluctuations from finite-amplitude Alfvén waves, though the relative importance of this mechanism in solar wind turbulence evolution remains debated \citep{Malara2000}.  

Most models of solar wind turbulence, and RDT specifically, assume a purely radial background magnetic field for simplicity. Beyond the Alfvén point, where the solar wind speed equals the Alfvén speed, the Sun’s rotation twists the field into the well-known Parker spiral (PS) (\citealt{Parker1958,Weber1967}). Close to the Sun, the spiral angle is very small, but beyond approximately $0.3$-$1$AU the azimuthal component becomes significant, such that it could alter wave propagation and nonlinear coupling \citep{Verdini2008,Owens2013, Verdini2018a}. 

The goal of this paper is to quantify how PS geometry modifies reflection-driven turbulence and the heating it causes in the super-Alfvénic solar wind. We argue that the relevant change to the standard \citet{Dmitruk2002} phenomenology of RDT due to the PS is that the evolution of the characteristic scales of the turbulence in the perpendicular and parallel directions are modified by the mean field's direction. We then study this phenomenology using the expanding-box model (EBM) \citep{Grappin1993} to follow a small plasma parcel advected outward, presenting a suite of three-dimensional compressible MHD simulations initialized with strongly outward-dominated $\bm{z}^+$ fluctuations. 
We find that the RDT picture remains broadly valid in PS geometry---the same reflection-driven dynamics operate, but the controlling scales change. Perpendicular expansion stretches eddies into pancake-like structures, while the growing azimuthal field in the PS rotates the mean field relative to those pancakes and therefore “cuts across” them; this geometric change produces smaller effective perpendicular scales than in the radial case. As a result the two transverse correlation lengths evolve differently and the effective perpendicular outer scale in a PS background tends to saturate rather than grow indefinitely. This causes the expansion-to-nonlinearity ratio, termed $\chi_{\rm exp}$, to decrease less rapidly and the system remains in a sustained, imbalanced nonlinear state. By contrast, in radial geometry the perpendicular scale grows with expansion, $\chi_{\rm exp}$ falls, and the cascade can shut off, leaving a magnetically dominated state with little associated heating.

The paper is organized as follows. In \cref{Theory} we develop the theoretical framework for turbulence evolution and introduce the compressible expanding-box magnetohydrodynamic (MHD) model employed in our study. We extend the simple reflection-driven phenomenology of \citet{Dmitruk2002} and show how PS geometry alters reflection, projected perpendicular scales, and the resulting nonlinear time-scales. In \cref{Numerics} we summarize the numerical method, list key parameters and resolution choices, and specify the initial conditions used in the simulation suite. In \cref{theoretical validation}, we test the theoretical expectations of \cref{Theory} by examining evolution of outward/inward energies, normalized cross-helicity, and the competing nonlinear and expansion timescales demonstrating how these diagnostics depend on the PS in general agreement with the phenomenology. \Cref{Observables} explores various other diagnostics for the purpose of comparing to observations, reporting the parametric evolution of cross-helicity and residual energy, switchback statistics, magnetic compressibility, and synthetic fly-by traces for comparison with spacecraft measurements. Finally, \cref{conclusion} summarizes the main results, discusses their implications for existing theory and in-situ observations. 
\section{Theoretical Framework}\label{Theory}
In this section, we build a theoretical model to describe how solar-wind expansion and large-scale magnetic field geometry shape Alfvénic turbulence. We introduce the expanding-box formalism to capture the kinematic effects of transverse expansion on fields and their length scales \citep{Grappin1993}, and formulate the dynamics in terms of Elsässer variables to make the reflection-driven coupling explicit. Building on this foundation, we outline reflection-driven turbulence and derive the \citet{Dmitruk2002} decay phenomenology, extending this to include the Parker spiral. This framework both motivates the simulations and diagnostics developed later and provides testable predictions for our simulations.
\subsection{Governing Equations for Reflection-Driven Turbulence}
The solar wind's turbulent evolution is governed by the interplay between reflection and nonlinear interactions. To model this in regions beyond the Alfvén critical point, where the flow becomes super-Alfvénic, we adopt the Expanding Box Model (EBM; \citealt{Grappin1993}), which approximates the spherical expansion of the solar wind in a Cartesian patch of wind co-moving with the constant radial flow velocity \( U \). Aligning the \( x \)-axis with the radial direction, the EBM introduces an expansion factor \( a(t) = R(t)/R_0 \), where \( R(t) = R_0 + U t \) is  heliocentric distance for some initial $R_0$, and modifies spatial gradients as \( \tilde{\nabla} = (\partial_x, a^{-1}\partial_y, a^{-1}\partial_z) \), viz., \({x}\) (radial) is unstretched, and \({y},{z}\) (azimuthal/normal) are stretched. $\dot{a} = U/R_0$ is the constant expansion rate due to $U$. The governing MHD equations in this frame become:

\begin{align}
    \frac{\partial \rho}{\partial t} + \tilde{\nabla}.(\rho \bm u) &= -2\frac{\dot{a}}{a}\rho, \label{eq:continuity} \\
    \frac{\partial \bm{u}}{\partial t} + (\bm{u} \cdot \tilde{\nabla}) \bm{u} &= -\frac{\tilde{\nabla} p}{\rho} + \frac{(\bm{B} \cdot \tilde{\nabla})\bm{B}}{4\pi\rho} - \frac{\dot{a}}{a} \mathbb{T} \cdot \bm{u}   + \mathcal{D}_{u}  , \label{eq:momentum} \\
    \frac{\partial \bm{B}}{\partial t} + (\bm{u} \cdot \tilde{\nabla}) \bm{B} &= (\bm{B} \cdot \tilde{\nabla}) \bm{u} - \frac{\dot{a}}{a} \mathbb{L} \cdot \bm{B}   + \mathcal{D}_{B} , \label{eq:induction}
\end{align}
\noindent where \( \mathbb{T} = \text{diag}(0,1,1) \) and \( \mathbb{L} = \text{diag}(2,1,1) \) encode anisotropic damping from angular momentum conservation and magnetic flux freezing, respectively (here diag denotes a diagonal matrix with the specified values), while $\mathcal{D}_{u}$ and $\mathcal{D}_{B}$ are the dissipative (viscous and resistive) terms that act only at small scales. In the real solar wind they correspond to kinetic damping processes, while in simulations they mainly represent explicit and numerical dissipation near the grid scale. 

We assume a locally isothermal equation of state, $p = c_s^2 \rho$, where $p$ is the thermal pressure, $c_s$ is the sound speed, and $\rho$ is the plasma density. By "locally isothermal," we mean that the temperature is spatially uniform throughout the computational domain at any given instant during the simulation. However, this does not imply that the temperature remains constant with heliocentric distance. Rather, we allow the temperature (and thus $c_s$) to evolve with $a$ in a manner that reproduces the cooling expected for expanding gas.
In particular, the sound speed varies with the expansion as $c_s \propto a^{-2/3}$, representing the cooling of the solar wind as it expands.
The locally isothermal approximation thus captures the background thermal evolution while simplifying the thermodynamics by enforcing spatial temperature uniformity at each time step. 

We take the $y$-axis to align with the ecliptic so that in the presence of Parker spiral the mean magnetic field has both radial ($x$) and azimuthal ($y$) components,
\({\bar{\bm B}} = {B_x} \hat{\bm x} + B_y\hat{\bm y}\). We denote spatial averages over the co-moving expanding domain by $\langle\cdot\rangle$ and fluctuations about a mean by $\delta(\cdot)\equiv (\cdot)-\langle(\cdot)\rangle$. In particular the mean magnetic field is $\bar{\bm B}=\langle\bm{B}\rangle,$ and we define the corresponding unit vector $\hat{\bm b}=\bar{\bm B}/\lvert{B}\rvert.$
From $\hat{\bm b}$ we form an orthonormal triad $(\hat e_\parallel,\hat e_T,\hat e_N$, see Fig.~\ref{fig:EMB_spiral_expansion})  by taking
\begin{align}
\hat e_\parallel &= \hat{\bm b}, \quad \hat e_T = \frac{\hat{\bm z}\times\hat{\bm b}}{\lvert\hat{\bm z}\times\hat{\bm b}\rvert}, \quad 
\hat e_N = \hat e_\parallel\times\hat e_T=\hat{\bm z}.
\end{align}
The magnetic-field components in this triad are
\begin{equation}
B_\parallel=\bm{B}\cdot\hat e_\parallel,\qquad 
B_T=\bm{B}\cdot\hat e_T,\qquad
B_N=\bm{B}\cdot\hat e_N.
\end{equation} \\
Due to expansion, the plasma mean quantities evolve with time. The magnetic field components follow $B_x = B_{x0}/a^2$, $B_y = B_{y0}/a$, and $B_z = B_{z0}/a$, and the density as \(\rho = \rho_0/a^2\), where, the subscript 0 is to refer to the values at $t = 0 (a = 1)$. The Alfvén velocity becomes,  
\[
\bm{v}_{\rm A} \equiv \frac{{\bar{\bm B}}}{\sqrt{4\pi\rho}} = \frac{v_{\rm A0,\textit{x}}}{a} \hat{\bm x} + {v}_{\rm A0,\textit{y}} \hat{\bm y}, \quad {v}_{\rm A0,\textit{x}} = \frac{B_{x0}}{\sqrt{4\pi\rho_0}}, \quad {v}_{\rm A0,\textit{y}} = \frac{B_{y0}}{\sqrt{4\pi\rho_0}},
\]  
such that the radial component \({v}_{\rm{A},\textit{x}} \propto 1/a\) decreases with expansion, while \({v}_{\rm{A},\textit{y}}\) remains constant. Based on these introduced scalings, the Parker angle $\Phi$, which quantifies the deviation of the background magnetic field from the radial direction in the $xy$-plane is,
\begin{equation}
\tan \Phi  = \frac{\langle{B}_y\rangle}{\langle{B}_x\rangle} \; =\; a \frac{{B}_{y0}}{{B}_{x0}}\propto a,
\end{equation}\label{eq:PS}indicating that if the initial background magnetic field has a non-zero transverse component, it will increasingly deviate from the radial direction as the system undergoes expansion. This is the manifestation of Parker spiral field in the local domain.

To develop our phenomenology, we assume locally constant density \( \rho \) and incompressible flow (\( \tilde{\nabla} \cdot \bm{u} = 0 \)). This allows us to simplify the equations and focus on the predominantly Alfvénic fluctuations. 
To study the evolution of these Alfvén waves—particularly the important phenomenon of reflection caused by gradients in the mean Alfvén speed---it is standard to reformulate the MHD equations in terms of the Elsässer variables: \( \bm{z}^\pm = \bm{u} \mp \bm{B}/\sqrt{4\pi\rho} = \bm{u} \mp \bm{b} \mp \bm{v_\mathrm{A}}\) (where $\bm{b} = \delta\bm{B}/\sqrt{4\pi\rho}$). The Elsässer formalism is useful because $\bm z^+$ and $\bm z^-$ signs distinguish fluctuations that propagate parallel and anti-parallel to the background field, respectively, clarifying the interplay of counter-propagating Alfvénic fluctuations. Combining the momentum equation~\eqref{eq:momentum} and the induction equation~\eqref{eq:induction} yields,

\begin{align}
\frac{\partial \bm{z}^\pm}{\partial t} \pm \left( \bm{v}_A \cdot \tilde{\nabla} \right) \bm{z}^\pm + \left( \bm{z}^\mp \cdot \tilde{\nabla} \right) \bm{z}^\pm + \tilde{\nabla} p'^\pm = -\frac{\dot{a}}{2a} \mathbb{T} \cdot \left( \bm{z}^+ + \bm{z}^- \right) \nonumber \\
- \frac{\dot{a}}{2a} \left( z_x^\pm - z_x^\mp \right) \hat{\bm x}   + \mathcal{D}^\pm  , \label{eq:final}
\end{align}

\noindent   where $\mathcal{D}^\pm$ represents the dissipative effects due to viscosity/resistivity that act on $\bm z^\pm$ , and $\tilde \nabla p'^\pm$ ensures $\nabla \cdot \bm{z}^\pm = 0$. We take the mean magnetic field $\bar{\bm B}$ to point in the positive radial direction, so that $\bm z^+$ perturbations propagate outward in the absence of reflection. 
  
\subsection{Wave-Action Elsässer Variables}
In an expanding medium like the solar wind, Alfvén wave amplitudes can naturally decay due to geometric WKB effects as well as true dissipation \citep{Heinemann1980}. To remove this trivial evolution, which manifests in Eq.~\eqref{eq:final} via $\bm z^\pm$ terms on right hand side, we rescale the Elsässer variables as \citep{Chandran2009,Meyrand2025},
 \begin{equation}
         \tilde{\bm z}^\pm \;=\;a^{1/2}\,\bm z^\pm \;\propto\;\frac{\bm z^\pm}{\sqrt{\omega_{\rm A}}},
 \end{equation}
where $\omega_{\rm A} = \bm{k} \cdot\bm{v}_{\rm A} = k_{x,0} (v_{\rm{A0},\textit{x}}/a) + (k_{y,0}/a) v_{\rm{A0},\textit{y}}  \propto 1/a$ is the local Alfvén frequency for wavenumber $\bm k$. This transformation isolates the physical processes---nonlinear cascade and reflection---that genuinely redistribute and dissipate wave energy. The nearly conserved quantity in this framework is the WKB wave-action density $|\bm z^\pm|^2/\omega_{\rm A}$, which remains constant without reflection or nonlinear coupling. Rewriting Eq.~\eqref{eq:final} in terms of $\tilde{\bm z}^\pm$ cancels all expansion–induced decay, leaving
\begin{equation}\label{eq:wave action}
\dot{a}\frac{\partial \tilde{\bm{z}}^\pm}{\partial a} \pm  \left( \bm{v}_A \cdot \tilde{\nabla} \right) \tilde{\bm{z}}^\pm + {a^{-1/2}} \left( \tilde{\bm{z}}^\mp \cdot \tilde{\nabla} \right) \tilde{\bm{z}}^\pm + \tilde{\nabla} \tilde{p}' = -\frac{\dot{a}}{2a} \tilde{\mathbb{T}}\cdot\tilde{\bm{z}}^\mp  + \mathcal{\tilde{D}}^\pm ,
\end{equation} where $\tilde{\mathbb{T}}= \rm{diag} (-1,1,1)$. 
Equation~\eqref{eq:wave action} shows explicitly how reflection seeds $\bm z^-$ from $\bm z^+$ fluctuations, while nonlinear interactions, which weaken in time due to the $a^{-1/2}$ factor, can sustain a cascade. The final term on the right-hand side of Eq.~\eqref{eq:wave action}, $-(\dot{a}/2a)\tilde{\mathbb{T}}\cdot\tilde{\bm{z}}^\mp$, represents an anisotropic reflection effect associated with the expansion. In the radial field case with perpendicular fluctuations, the $\tilde z^\pm_x$ is negligible and the dominant reflection occurs through $\tilde z_{y,z}^- \propto -\tilde z_{y,z}^+$, thereby naturally causing $\langle \tilde{\bm{z}}^+ \cdot \tilde{\bm{z}}^- \rangle <0$; where $\langle \tilde{\bm{z}}^+ \cdot \tilde{\bm{z}}^- \rangle/2 = (\langle \bm u^2\rangle-\langle \bm b^2\rangle)/2$ is the wave-action residual energy $\tilde E_{r}$. However, with either large-amplitude spherically polarized fluctuations or an oblique background field, there will be significant $z_x^\pm$ component, so that we also have contribution from $\tilde z_x^- \propto \tilde z_x^+$. This anisotropic coupling could thus moderate the growth of the cross correlation $\langle \tilde{\bm{z}}^+ \cdot \tilde{\bm{z}}^- \rangle$, resulting in systematically smaller values of $\lvert \langle \tilde{\bm{z}}^+ \cdot \tilde{\bm{z}}^- \rangle\rvert$ in the PS configuration compared to the radial case. Physically, this implies that strong negative correlations between counter-propagating Elsässer fields are expected in the radial case, whereas in the PS geometry the anisotropic reflection term reduces the magnitude of $\langle \tilde{\bm{z}}^+ \cdot \tilde{\bm{z}}^- \rangle$.

\subsection{Energy Dynamics in Expanding Solar Wind Turbulence} \label{energy dynamics}
In MHD turbulence, energy is distributed between kinetic and magnetic fluctuations, with nonlinear interactions transferring energy across scales. 
The wave-action Elsässer energy densities are defined by $\tilde{E}^\pm \equiv \langle |\tilde{\bm{z}}^\pm|^2 \rangle / 4$. Unlike homogeneous MHD these energies are not ideally conserved under solar wind expansion, because reflection acts as either a source or sink depending on the correlation between counter-propagating modes.
Multiplying the evolution Eq.~\eqref{eq:wave action} by $\tilde{\bm{z}}^\pm$, then averaging gives,
\begin{equation}\label{energy_eq}
    \dot{a} \frac{\partial}{\partial a} {\tilde{E}^\pm} = -\frac{\dot{a}}{4a} { \langle \tilde{\bm{z}}^+ \cdot \mathbb{T} \cdot \tilde{\bm{z}}^- \rangle} + \frac{\dot{a}}{4a} \langle \tilde{z}_x^+ \tilde{z}_x^- \rangle +   \mathcal{\tilde{D}}^\pm  ,
\end{equation}  
because we expect predominantly Alfvénic fluctuations, which are perpendicular to $\bar{\bm B}$, in the case of the purely radial field the term $\langle \tilde{z}_x^+ \tilde{z}_x^- \rangle$ is likely negligible, so the energy evolution simplifies to $\dot{a} \, \partial \tilde{E}/\partial a = -(\dot{a}/2a) \tilde E_{r}$ \citep{Meyrand2025}. The reflection thus acts as a source or sink of wave-action energy depending on the sign of the residual energy $\tilde E_{r}$ (note that any change in fluctuation energy is balanced by the background-flow energy budget; \citealt{Chandran_2015}).
In the absence of dissipation and external forcing, $\tilde{E}^+ - \tilde{E}^-$ is conserved by the combined dynamics of wave reflection and advection, because the corresponding source terms are identical in the $+$ and $-$ equations and cancel upon subtraction; making this difference the true wave-action invariant \citep{Heinemann1980}.
 
In the Parker spiral configuration, the $+ \langle \tilde{z}_x^+ \tilde{z}_x^- \rangle$ term could become significant even when $\bm z^\pm$ is predominantly perpendicular to $\bm B$, and we likewise see from Eq.~\eqref{eq:wave action} that it is naturally driven with the correlation required to act as a source of $\tilde{E}^\pm$.

\subsection{Reflection-driven Turbulence Phenomenology}\label{Phenomnology}
The RDT phenomenology involves the consideration of different balance of key terms. It is helpful to sketch Eq.~\eqref{eq:wave action} as
\begin{equation}\label{eq:cartoon}
    \dot{a}\frac{\partial \tilde{\bm{z}}^\pm}{\partial a} + \mathcal{L}^\pm  \tilde{\bm{z}}^\pm + \mathcal{N}^\pm \tilde{\bm{z}}^\pm = -  \mathcal{R}^\pm \tilde{\bm{z}}^\mp,
\end{equation}
where $\mathcal{L}^\pm = \bm{v}_A \cdot \tilde \nabla$ represents Alfvénic propagation, $\mathcal{N}^\pm = {\bm{z}}^\mp \cdot \tilde{\nabla} = a^{-1/2} \tilde{\bm z}^\mp \cdot \tilde{\nabla} \sim {z}^\mp/\ell_\perp$ represents nonlinear terms that will damp out ${z}^\pm$ (here $\ell_\perp$ denotes the perpendicular correlation length scale of fluctuations and ${z}^+$ denotes the characteristic outer-scale amplitude of outward-propagating fluctuations at scale $\ell_\perp$), and reflection is $\mathcal{R}^\pm = -\dot{a} / 2 a \mathbb{\tilde T}$. 
The characteristic timescales associated with the terms in Eq.~\eqref{eq:cartoon} can differ in their scaling.
The Alfvén propagation time along the mean field is $\tau_{\rm A} = \ell_\parallel/v_{\rm A}$ ($\ell_\parallel$ is the parallel correlation length), while the nonlinear interaction time across the field is $\tau_{\rm nl} = \ell_\perp/{z}^\mp$. Both $\tau_{\rm A}$ and $\tau_{\rm nl}$ decrease at smaller scales, since they depend directly on the characteristic sizes of turbulent structures along and across the mean magnetic field respectively. In contrast, the expansion time $\tau_{\rm exp}= a/\dot{a}$ is independent of scale.
Following \cite{Meyrand2025}, we define the dimensionless ratios;
\begin{equation}
    \chi_{\rm exp} = \frac {\mathcal{N} }{\mathcal{R}}  = \frac{\tau_{\rm exp}}{\tau_{\rm nl}}, \quad \chi_{\rm A} = \frac {\mathcal{N} }{\mathcal{L}} = \frac{\tau_{\rm A}}{\tau_{\rm nl}},
\end{equation}which quantify the relative importance of expansion and Alfvén propagation compared to nonlinear interactions \citep{GS95}. These ratios will play a key role in controlling the dynamics of the turbulent cascade in an expanding solar wind.

The standard phenomenology \citep{Dmitruk2002,Verdini2007,Chandran2009} balances these terms to model turbulence and heating, making two key balancing assumptions. First, for $\tilde{z}^-$ (backward waves) propagation and time variation are assumed negligible, the reflection source ($\mathcal{R}^- \tilde{z}^+$) balances the nonlinear sink ($\mathcal{N}^- \tilde{z}^-$):
\begin{equation}
\mathcal{N}^- \tilde{z}^- \sim \mathcal{R}^- \tilde{z}^+\implies  \tilde{z}^- \sim  \frac{\mathcal{R}^- \tilde{z}^+}{\mathcal{N}^-} \sim \frac{\dot{a}}{2a}\frac{\tilde{z}^+}{({ z}^+/\ell_\perp)} = a^{1/2}\left(\frac{\dot{a}}{2a}\right) \ell_\perp,
\end{equation}
and using $(\dot a/a)/(z^+ /\ell_\perp)=\chi_{\rm exp}^{-1}$ gives,
\begin{equation}
    \tilde{z}^- \simeq \frac{\tilde{z}^+}{\chi_{\rm exp}}\;  \Rightarrow \; \frac{{z}^-}{{z}^+} \simeq \chi_{\rm exp}^{-1} .
\end{equation}\label{RDT Prediction}This means that the $\tilde{z}^-$ amplitude is small---i.e., turbulence is highly imbalanced---when reflection is weak or nonlinearity is strong $\chi_{\rm exp} \gg 1$. The second assumption concerns the decay of $\tilde{\bm z}^+$. In the $\tilde z^+$ equation we neglect reflection because $\tilde z^-\ll\tilde z^+$. The linear propagation terms ($\mathcal L^\pm\tilde{\bm z}^\pm$) doesn’t directly lead to decay---it merely transports fluctuations along the mean field---so $\partial\tilde{\bm z}^+/\partial t$ is dominated by damping at the nonlinear rate ($\mathcal N^+$): 
\begin{equation}\label{energy_prediction}
    \dot{a}\frac{\partial \tilde{{z}}^+}{\partial a} \simeq - \mathcal{N}^+ \tilde{{z}}^+ \implies \dot{a}\frac{\partial}{\partial a}\tilde{{z}}^+ \simeq - \frac{{ z}^-}{\ell_\perp}\tilde{{z}}^+ \simeq -\frac{\dot{a}}{2 a} \tilde{z}^+  \implies \tilde{z}^+ \propto a^{-1/2}
\end{equation}
This corresponds to $z^+\propto a^{-1}$ or (in radial field) $z^+/v_{\rm A} \sim \text{const}$. 
The heating rate $\mathcal{Q}$, which is the rate at which energy in the outward-propagating mode $\tilde{z}^+$ is damped, can be calculated using Eq.~\eqref{energy_eq} by multiplying with $\rho V$ to get actual energy (see, \citealt{Perez2021})\footnote{From Eq.~2.12 of \cite{Perez2021}, neglecting the residual energy and $E^-$, then assuming super-Alfvénic flow with $U=\mathrm{const}$ and $\hat{\bm b}=\hat{\bm r}$ (where $\hat{\bm r}$ is the radial unit vector in spherical coordinates), the steady energy balance reduces to $\nabla\cdot(\hat{\bm r}\,U\,\rho\,E^+)+\tfrac12\,\rho E^+ U\,\nabla\cdot\hat{\bm r}=-Q$. Using $\nabla\cdot\hat{\bm r}=2/r$, this becomes $\partial_r(r^2 \rho U E^+)/r^2+\rho U E^+/r=-{Q}$. With $U=\mathrm{const}$ and $\rho r^2=\rho_0$, we get $U\rho_0\big(\tfrac{1}{r^2}\partial_r E^+ + \tfrac{1}{r^3}E^+\big)=-\mathcal{Q}$. Since $E^+=r\tilde E^+$, we obtain $U\rho_0\big[\tfrac{1}{r^2}\partial_r (\tilde E^+/r) + \tfrac{1}{r^4} \tilde E^+\big]= U\rho_0/{r^3}\partial_r \tilde E^+ =-{Q}$. This matches Eq.~\eqref{heating_rate}, if we note that the extra $1/r^2$ arises because this is heating per unit volume, whereas Eq.~\eqref{energy_prediction} is heating per unit co-moving volume.}
, where $ V \propto a^2$ is the volume of the domain. This gives
\begin{equation}\label{heating_rate}
    \mathcal{Q}= \frac{\dot{a}}{a} \frac{\partial}{\partial a} {(\rho \tilde{E}^+ a^2)} = \rho_0\frac{\dot{a}}{a} \frac{\partial}{\partial a} \frac{\langle|\tilde{\bm z}^+|\rangle^2}{4} \approx 
    \rho_0 \frac{\dot{a} \tilde{z}^+}{2a}  \frac{\partial \tilde{z}^+}{\partial a} \approx \rho_0\frac{\dot{a}}{4 a^2} (\tilde{z}^+)^2,
\end{equation}
where we neglected the residual energy and use $\langle|\tilde{\bm z}^+|\rangle^2 \approx (\tilde{z}^+)^2$ (mean square amplitude). The phenomenology in \cref{energy dynamics} predicts $\tilde{E}\propto a^{-1}$, which implies a heating rate per co-moving volume $\mathcal{Q}\propto a^{-3}$. Converting to physical units (physical volume $\propto a^2$) gives a heating rate per physical volume $\mathcal{\tilde{Q}}\propto a^{-5}$. 
  The heating rate $\mathcal{Q}$ represents the actual dissipation of turbulent energy into thermal energy through viscous and resistive processes, not the reversible redistribution of energy between outward and inward propagating modes. Reflection converts $\bm z^+$ into $\bm z^-$ and vice versa, and the nonlinear turbulent cascade transfers energy from large scales to small scales, but only when the cascade reaches dissipative scales is energy irreversibly removed and counted in $\mathcal{Q}$.
In our simulations, explicit viscosity and resistivity are not included; instead, numerical dissipation at the grid scale in Athena++ serves as an effective dissipation mechanism that captures the cascade. The rate $\mathcal{Q}$ therefore measures how rapidly the outward energy $\tilde{E}^+$ is depleted by this dissipative process, which is fundamentally distinct from the reversible energy exchange between $\bm z^+$ and $\bm z^-$ driven by reflection.
We note that while kinetic effects likely dominate dissipation in the actual solar wind plasma, our MHD framework with numerical dissipation at the grid scale captures the essential phenomenology of energy removal at small scales.
 
Interestingly, this heating rate does not depend on the nonlinear rate or scale of the turbulence, because a smaller transverse scale $z^+$ damps $z^-$ more rapidly, a smaller $\ell_\perp$ implies that $z^-$ has smaller amplitude for the same reflection rate $\mathcal{R}$. This lower amplitude then cancels the increased nonlinear rate coming from the smaller scale of $z^-$, thus leading to the same overall efficiency at damping $z^+$.

The RDT framework rests on following assumptions that demand careful scrutiny. First, the nonlinear damping rate provided by the turbulence is  \( {z}^\mp/\ell_\perp \), implying strong turbulence (\(\chi_{\rm A} \sim 1\)) for validity (but see related weak turbulence version in \citealt{Chandran2019}), which leads to the same scaling. Second, the phenomenology assumes that a single dominant scale \(\ell_\perp\) governs both \(\tilde{z}^+\) and \(\tilde{z}^-\); this $\ell_\perp$ cancels in the heating rate \(\mathcal{Q}\), but in principle could be different for \(\tilde{z}^+\) and \(\tilde{z}^-\) (indeed, this was seen by \citealt{Meyrand2025}). Third, anomalous coherence, with $\tilde{z}^-$ carried along by $ \tilde{z}^+ $, is required to preserve the $\tilde{z}^-$ balance and to justify neglecting the ($ v_{\rm A} \cdot \nabla $) term in our first assumption. Fourth, \(\tilde{z}^-\) must adjust to reflection-nonlinear equilibrium faster than background changes; this requirement reduces to \(\chi_{\rm exp} \gtrsim 1\). Fifth, reflection in the \(\tilde{z}^+\) equation must be negligible (\(\mathcal{R}^+ \tilde{z}^- \ll \mathcal{N}^+ \tilde{z}^+\)), which likewise reduces to the same \(\chi_{\rm exp} \gtrsim 1\) condition. 

Thus the phenomenology is self-consistent only for sufficiently large amplitude $\tilde z^+$ such that $\chi_{\rm exp}\gg1$, which implies strong imbalance $\tilde z^-\ll\tilde z^+$. We therefore expect the system to transition to the balanced regime once the outward Elsässer energy decays to the point that $\chi_{\rm exp}\sim1$. Conversely, when $\chi_{\rm exp} < 1$, the ordering breaks down, reflection becomes dominant and the dynamics from \eqref{heating_rate} likely revert toward reflection-dominated (essentially linear) behavior. In this regime with a radial $\bar{\bm B}$, the system evolves into a magnetically dominated, balanced state in which $\tilde{\bm z}^+ \simeq -\tilde{\bm z}^-$, and $\tilde{E}$ starts growing in time (see Eq.~\ref{energy_eq}). Nonlinear dissipation is strongly suppressed, so the turbulent heating effectively shuts off \citep{Meyrand2025}.

\subsubsection{The role of Parker spiral}\label{sub:PS Theory}
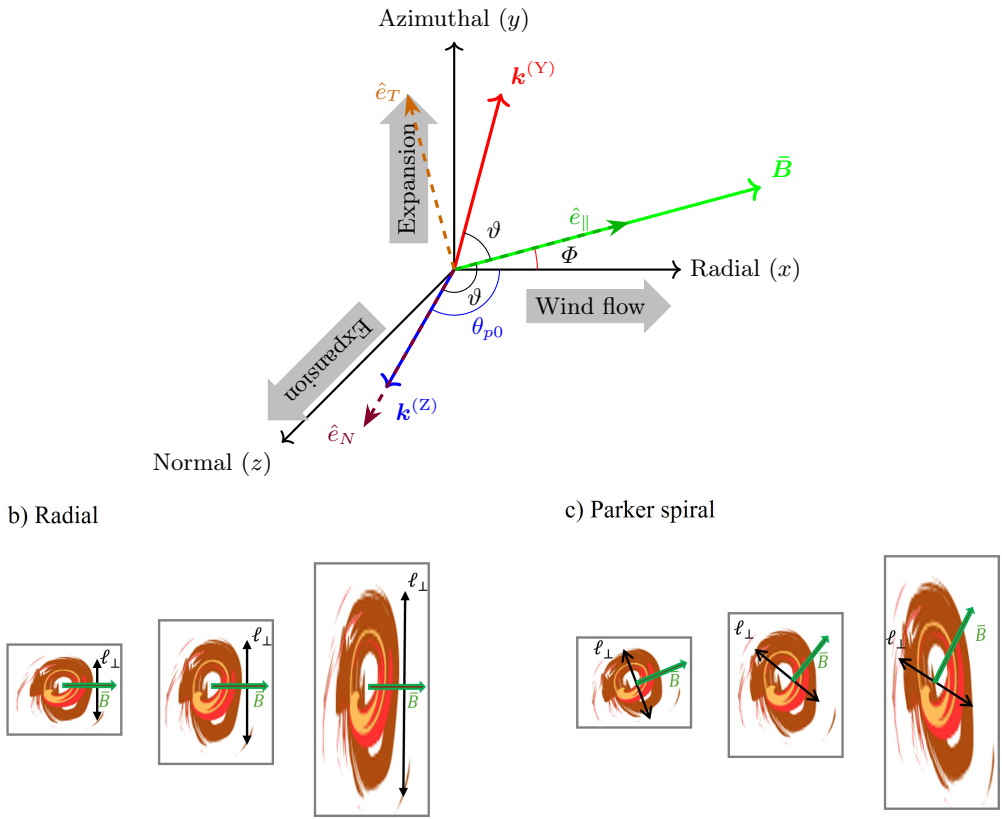
\begin{figure}
\centering
\begin{tikzpicture}[scale=1.2, every node/.style={font=\small}]
\begin{scope}[shift={(2.5,3.5)}] 
  \draw[->, thick] (0,0) --++(2.5,0) node[right] {Radial $(x)$};
  \draw[->, thick] (0,0) --++(0,2.5) node[above] {Azimuthal $(y)$};
  \draw[->, thick] (0,0) --++(-1.9,-1.9) node[below left] {Normal $(z)$};
  \draw[->, very thick, green] (0,0) --++({3.5*cos(-15)},{3.5*sin(-15)}) node[right] {$\bm{\bar{B}}$};
  \path 
      (4.5,0) coordinate (R)   
      (0,0)   coordinate (O)   
      ({2*cos(-15)},{2*sin(-15)}) coordinate (S);  
  \draw pic[
    draw=red,
    angle eccentricity=1.4,
    angle radius=1.1cm,
    "$\Phi$"
  ] {angle = S--O--R};
  \draw[->, very thick, red] (0,0) --++({2*cos(75)},{2*sin(75)}) node[above right] {$\bm{k}^{\rm (Y)}$};
  \path 
    ({3.5*cos(-15)},{3.5*sin(-15)}) coordinate (Btip)   
    (0,0)   coordinate (O)   
    ({3.5*cos(75)},{3.5*sin(75)}) coordinate (S2);  
  \draw pic[
    draw=black,
    angle eccentricity=1.5,
    angle radius=0.5cm,
    "$\vartheta$"
  ] {angle = Btip--O--S2};
  \coordinate (B) at ({3.5*cos(-15)},{3.5*sin(-15)}); 
  \coordinate (o) at (0,0);
  \coordinate (K) at ({1.5*cos(120)},{-1.5*sin(120)}); 
  \draw[->, very thick, blue] (O) -- (K) node[below right] {$\bm{k}^{\rm (Z)}$};
  \draw pic[
    draw=black,
    angle radius=0.3cm,
    angle eccentricity=1.5,
    "$\vartheta$"
  ] {angle = K--o--B};
  \draw pic[
    draw=blue,
    angle radius=0.6cm,
    angle eccentricity=1.5,
    "$\color{blue}{\theta_{p0}}$"
  ] {angle = K--o--R};
  \node[single arrow, fill=gray!50, minimum height=1.0cm, 
        single arrow head extend=1ex, rotate=90] at (-0.5,1.0) {Expansion};
  \node[single arrow, fill=gray!50, minimum height=0.9cm, 
        single arrow head extend=1ex, rotate=225] at (-1.4,-1.0) {Expansion};
  \node[single arrow, fill=gray!50, minimum height=0.8cm, 
        single arrow head extend=1ex, rotate=0] at (1.5, 0.4) {Wind flow};
  \def\L{2.0}   
  \def\LN{2.0}  
  \draw[->, dashed, very thick, green!70!black, >=Stealth] 
    (O) -- ({\L*cos(-15)},{\L*sin(-15)}) 
    node[pos=0.7, below right] {$\hat{e}_\parallel$};
  \draw[->, dashed, very thick, orange!80!black, >=Stealth]
    (O) -- ({\L*cos(75)},{\L*sin(75)}) node[pos=0.90, above left] {$\hat{e}_T$};
  \draw[->, dashed, very thick, purple!70!black, >=Stealth]
    (O) -- ({\LN*cos(-120)},{\LN*sin(-120)}) node[pos=0.90, below left, xshift=-2pt, yshift=-1pt] {$\hat{e}_N$};
\end{scope}
\end{tikzpicture}
\includegraphics[width=\linewidth]{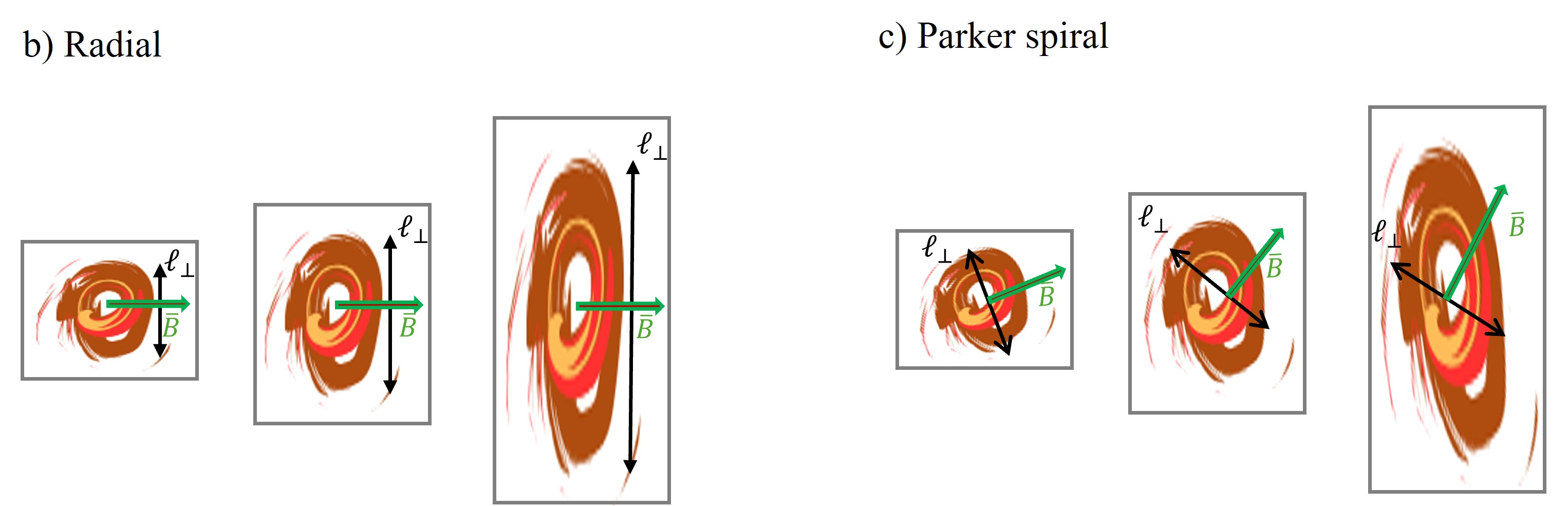}
\caption{ Schematic of an expanding plasma parcel in the solar wind. (Top) Geometry of the expanding box with axes aligned to radial ($x$), azimuthal ($y$), and normal ($z$) directions. The solar wind initially flows radially outward, while the box expands transversely. The mean magnetic field $\bar{\bm B}$ (green) is radial near the Sun but rotates into a Parker spiral with angle $\Phi$ as $a(t)$ increases, so that $B_y/B_x \sim a(t)$. Wave-vectors $\bm{k}$ are shown at angle $\vartheta$ to $\bar{\bm B}$, with azimuthal ($k^{(Y)}$) and normal ($k^{(Z)}$) components indicated. (Bottom) Comparison of eddy evolution under (a) purely radial and (b) Parker-spiral expansion. In the radial case, the perpendicular scale $\ell_\perp$ grows uniformly with expansion. In the PS case, rotation of the mean field changes $\ell_\perp$ creating 3-D anisotropy of the eddies.}
\label{fig:EMB_spiral_expansion}
\end{figure}

The key new contribution of this work is the extension of RDT phenomenology to the Parker spiral geometry. We argue that the underlying dynamics remain similar to those in radial field, but that the spiral geometry introduces a key geometric modification affecting the evolution of eddy shapes. 
In a purely radial field, as the solar wind expands, eddies are stretched by expansion preferentially transverse to the magnetic-field direction, becoming increasingly pancake-shaped: $\ell_\perp$ increases while $k_\perp$ decreases. This stretching rapidly reduces the nonlinear turnover rate, accelerating the approach to $\chi_{\rm exp}= 1$. However, as the Parker spiral twists the mean magnetic field away from purely radial, it effectively rotates the mean field relative to how eddy structures are stretched by expansion. This rotation causes the field to “cut across” different parts of the eddies, so the eddies no longer remain perfect pancakes with respect to the local mean field and develop three-dimensional, anisotropic outer-scale shapes. A cartoon of this physics is shown in Fig.~\ref{fig:EMB_spiral_expansion}. To capture this anisotropic deformation, we use the local orthogonal basis $(\hat{e}_\parallel, \hat{e}_T, \hat{e}_T)$ to define two transverse outer scales, $\ell_{\perp,\rm T}$ and $\ell_{\perp,\rm N}$, characterizing the turbulent eddy scales along $\hat{e}_T$ and $\hat{e}_N$. The nonlinear interaction time is then expected to be determined by the smaller of these two transverse scales:
\begin{equation}
    \tau^\pm_{\rm nl} \sim \frac{\min (\ell_{\perp,\rm T},\; \ell_{\perp,\rm N})}{{z}^\mp}
\end{equation}
As $\ell_\perp(a)$ starts flattening or even decreases with $a$, this decreases the outer scales nonlinear timescale $\tau_{\rm nl}$. Consequently, this helps maintain $\chi_{\rm exp}$ large enough to prolong the imbalanced cascade. In this way, the competition between rotation and expansion means that the Parker spiral geometry prevents the eddies from becoming indefinitely flattened into pancakes, sustaining turbulence and heating farther out in the solar wind. 

To quantify that intuition, we will calculate---under the simplifying assumption that outer scales evolve linearly with the background flow---how the wave obliquity, projected wave vector angles, perpendicular wavenumber $k_\perp$, nonlinear time $\tau_{\rm nl}$, and thus $\chi_{\rm exp}$ evolve with expansion. For the evolution of $z^+$, we adopt the same phenomenology discussed above, with $\tilde{z}^+\propto a^{-1/2}$. This is motivated by the fact that the reflection, although differing in sign in the $x$-direction, has the same magnitude in all directions. Related effects were previously explored in the context of switchbacks by \cite{Squire2022}. It is helpful to define the following angles: the wave obliquity parameter $\vartheta \approx \arccos{(\hat{\bm{k}} \cdot \bar{\bm B})}$ quantifies the angle between the wave vector direction $\hat{\bm{k}}$ and the mean magnetic field ${\bar{\bm B}}$, which determines $\ell_\perp$; $\theta_{p0}$ measures the initial angle between the wave vector and the radial direction; and \(\varphi\) denotes the angle of the wave vector projected onto the tangential-normal ($y$-$z$) plane, such that \(\varphi = 0\) corresponds to \(\bm{k}\) aligned with the tangential direction and \(\varphi = \pi/2\) to \(\bm{k}\) aligned with the normal direction (Fig.~\ref{fig:EMB_spiral_expansion}). 

\begin{figure}
    \centering
    \includegraphics[width=\linewidth]{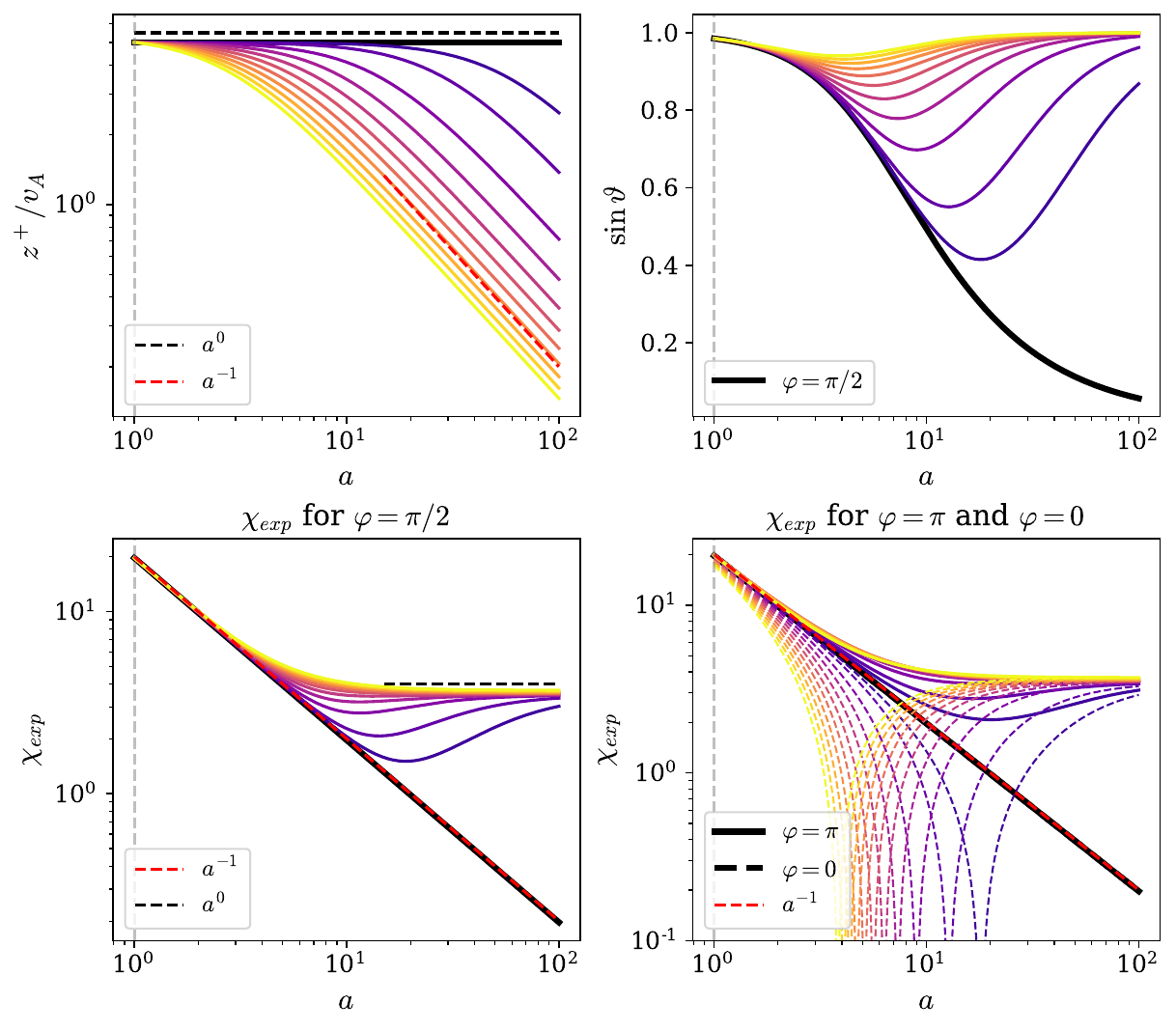}
    \caption{
    Evolution of key parameters for waves under solar-wind expansion, plotted versus expansion factor $a$.  Thick black: radial field ($\Phi_0=0^\circ$).  {Colored:} Parker spiral cases ($\Phi_0=2^\circ\text{--}20^\circ$, darker colors correspond to  {smaller}   $\Phi_0$). (a) Wave amplitude $z^+/v_{\rm A}$ stays roughly constant for the purely radial case, and in the Parker spiral (PS) case remains constant initially but then decays as $\propto a^{-1}$ once the azimuthal component becomes significant and $v_{\rm A}\propto a^0$. (b) Obliquity $\sin\vartheta(a)$: the angle between $\bm k$ and $\bar{\bm B}$ initially decreases and then increases again as the mean field rotates azimuthally, producing a clear inflection with a PS. (c) Expansion–cascade parameter $\chi_{\rm exp} = (k_\perp\,z^+_\perp)/(\dot a/a)$ for the out-of-plane case ($\varphi = \pi/2$): turbulence is sustained while $\chi_{\rm exp} \gtrsim 1$; $\chi_{\rm exp}$ initially decays as $\propto a^{-1}$ for both radial and Parker spiral cases, but for PS it starts increasing and eventually flattens as $a^{0}$ once the azimuthal component becomes dominant. (d) $\chi_{\rm exp}$ for in-plane wave vectors: shown for $\varphi=\pi$ (solid) and $\varphi=0$ (dashed); in both cases, the wave vector lies in the $x$--$y$ plane. For $\varphi=0$, the wave passes through purely parallel propagation.}
    \label{fig:Theory}
\end{figure}
As shown in Fig.~\ref{fig:Theory}, for the purely radial magnetic field case ($\Phi_0=0$), the system expands such that the eddy scale grows ($\ell_\perp \propto a$), the fluctuation amplitude decays ($z^+ \propto a^{-1}$), and $v_{\rm A}\propto a^{-1}$; the ratio $ z^+/v_{\rm A}$ is thus constant with expansion. In the PS case the growing azimuthal component causes $v_{\rm A}$ to decrease more slowly (and eventually become nearly constant $v_{\rm A} \propto a^{0}$), so $z^+/v_{\rm A}$ first remains constant and then decreases approximately as $a^{-1}$ once the spiral angle becomes large. The non-linear time scale increases because $k_\perp \propto a^{-1}$, giving $k_\perp z^+ \propto a^{-2}$ ($\tau_{\rm nl} \propto a^2$). This implies $\chi_{\rm exp}$ also decays as $\chi_{\rm exp} \propto a^{-1}$, rapidly approaching $\chi_{\rm exp}\leq 1$, where the phenomenology breaks.

In the Parker spiral case, $\sin \vartheta$ initially decreases as in the radial case, but as $a$ approaches $a_{\min} = \sqrt{\cot \theta_{p0} \,\cot \Phi_0}$, its evolution deviates: $\sin \vartheta$ reaches a minimum and then increases again as $\overline{\bm B}$ rotates toward the perpendicular direction (see Fig.~\ref{fig:Theory}b and also \citealt{Squire2022}). 
While the detailed dynamics are complex and nonlinear, we assume the overall RDT framework still applies such that $z^+\propto a^{-1}$, but the evolving geometry changes the scaling of $k_\perp$. Initially $k_\perp$ decays with expansion as in the radial case, but as the mean field rotates relative to the eddies, making wave vectors oblique and driving $k_\perp$ back up, thereby increasing the non-linear turnover rate. As a result, instead of continuously decaying, $\chi_{\rm exp}$ levels off staying roughly constant ($\propto a^0$) or even growing at intermediate times, delaying indefinitely the turbulence shutoff that occurs at $\chi_{\rm exp} < 1$ and maintaining the high imbalance (see Fig.~\ref{fig:Theory}). 

This revival is complicated by wave orientation in the $\hat{e}_T$-$\hat{e}_N$ plane. When wave vectors lie perpendicular to the spiral plane $(\varphi= \pi/2, \; p^{(Z)})$, $\chi_{\rm exp}$ exhibits a robust rebound. For in-plane orientations $(p^{(Y)})$, however, waves can pass through a purely parallel propagation phase $(\vartheta \to 0)$ suggesting a temporary collapse of nonlinear interactions before partial recovery. However, this is likely not significant: it occurs only when the wave vectors lie perfectly in the plane of the Parker spiral ($\varphi=0$), which is a very special case. As discussed above, the Parker spiral geometry squeezes eddies into 3D anisotropic structures with the shortest perpendicular scale likely determining the nonlinear turnover time. Thus, even $\ell_\perp$ becomes large in one direction the cascade can presumably continue. Note that an important caveat about the assumption $\ell_y=\ell_z \propto a$ may not hold strictly, since MHD nonlinear interactions tend to elongate structures preferentially along the local field direction $\hat{e}_\parallel$, thereby presumably changing the perpendicular growth for the linear evolution discussed above.

We now test the above ideas with a controlled numerical experiment suite. Using compressible expanding-box MHD simulations, we evolve initially outward $z^+$ fields and measure the properties introduced above to validate the phenomenology and quantify its parameter dependence. The following sections present these numerical results, compare them directly with the theoretical expectations, and highlight where the simulations confirm, refine, or challenge our simple model.

\section{Numerical Methods and Simulation setup}\label{Numerics}
\subsection{Expanding box model numerical}
To solve the EBM Eqs.~\eqref{eq:continuity}--\eqref{eq:induction}, we use the finite-volume astrophysical code Athena++ \citep{Stone2020}. We employ the Harten-Lax-van Leer Discontinuities (HLLD) Riemann solver \citep{Mignone2007}, modified to include expansion effects. To simplify the expansion terms and enhance numerical stability, we transform variables as follows \citep{Johnston2022}:
\begin{equation}\label{var}
    \rho = \lambda^{-1} \rho', \quad \bm{u} = \Lambda \bm{u}', \quad \bm{B} = \lambda^{-1} \Lambda \bm{B}', \quad \nabla' = \Lambda \tilde{\nabla},
\end{equation}
where, $\Lambda=\mathrm{diag}(1,a,a)$ and $\lambda=a^2$. 
Following this variable change and expressing the EBM MHD equations in conservative form, we obtain,
\begin{equation}\label{3.1}
    \frac{\partial \rho'}{\partial {t}} + {\nabla}' \cdot (\rho' \bm{u}') = 0
\end{equation}

\begin{equation}\label{3.2}
\frac{\partial \left(\rho'\bm{u}'\right)}{\partial t} + {\nabla}' \cdot\left(\rho' \bm{u}' \bm{u}' + \left(p+\frac{B^2}{2}\right) \Lambda^{-2} - \frac{\bm{B}' \bm{B}'}{\lambda}\right) = -\left( \frac{d \ln{\lambda}}{dt}\right) \mathbb{T}'.\rho'\bm{u'} 
\end{equation}

\begin{equation}\label{3.3}
    \frac{\partial \bm{B}'}{\partial {t}}-\nabla'\times \left(\bm{u}'\times\bm{B}'\right)= 0
\end{equation}
These Eqs.~\eqref{3.1}--\eqref{3.3} are quite similar to the ideal MHD equations, since the continuity 
and induction equations have no expansion or source terms. Instead, the momentum equation and the variable defined in Eq.~\eqref{var}, now incorporates all 
the effects of expansion.

\subsection{Simulations setup and initial conditions}
Numerical details of all simulations are given in Table~\ref{tab1:sim_params}. 
All simulations employ periodic boundary conditions in $x,y,z$, as is standard for local EBM studies. This is justified because the computational patch remains small compared with the heliocentric distance, allowing the EBM mapping of local spherical expansion onto a Cartesian, periodic domain. Periodicity reduces computational cost and therefore enables higher numerical resolution to capture the turbulent cascade across a broad range of scales. We stress the limitations: periodic domains omit open-boundary effects (global inflow/outflow) and can box-limit the longest parallel modes (affecting $\ell_\parallel$ and near-2D structures). To mitigate these effects we use an initially anisotropic (elongated) domain so transverse stretching is naturally represented as $L_y,L_z\propto a$. Within these bounds the model effectively captures the local, expansion-driven dynamics of reflection-driven turbulence beyond the Alfvén critical point.

We set the initial box dimensions to $L_{x0}=1$ and $L_{y0}=L_{z0}=0.2$.
We use the expansion factor $a$ to parameterize the evolution of the system, where $a$ is directly related to heliocentric distance via $a=R/R_0$, with $R_0$ being the radial distance at which the simulation is initialized ($a=1$). The choice of $R_0$ is arbitrary in the sense that it sets the reference scale and simply rescales other quantities (e.g., $v_{\rm A}$) without changing the dimensionless physics. To aid interpretation and correspondence with observational regimes, we can map $a$ to actual solar distances using the evolving Parker spiral angle $\Phi(a)$. Assuming a Parker spiral angle of $45^\circ$ occurs at 1 AU \citep{Borovsky2010}. In our simulations with initial Parker spiral angle $\Phi_0=2^\circ$, the value $a\, \approx \,28.64$ corresponds to 1 AU, while for $\Phi_0=5^\circ$, we have $a\approx11.5$ at 1 AU.
Importantly, the mapping between the heliocentric distance and PS is not unique: it depends on the PS calibration (e.g. the reference angle at 1\,AU), the latitudes, and on how one chooses the simulation inner boundary.
At high latitudes, the PS is much weaker, so off-ecliptic regions remain nearly radial even out to several AU. Note that, scanning over $\Phi_0$ at fixed initial conditions is best interpreted as starting from the same initial $\chi_{\rm exp,0}$ (discussed below) at different heliocentric distances. 
 
As the system expands, the box elongates in the transverse directions; by $a=50$, the domain becomes strongly oblate, resembling a pancake-like structure. At $a=1$, we initialize the background magnetic field ${\bar{\bm B}}$ in the $xy$ plane with $|{\bar{\bm B}}|=1$ and $B_{y,0} < 0$, defining the initial Parker‐spiral angle $\Phi_0$.

Waves exhibiting nearly constant magnetic-field strength and strong Alfvénic correlation $(\delta \bm B_\perp \propto \delta \bm u_\perp)$ are observed in the solar wind. These waves are called spherically polarized because the constant-$|\bm B|$ is preserved in the fluctuations. This property can be quantified through
\begin{equation}
\label{eq:CB2}
C_{B^2} \equiv \frac{\delta\left(| B|^2\right)}{\left(\delta\bm{B}\right)^2}={\frac{\left({{\left\langle\left(B^2 - \langle B^2\rangle\right)^2\right\rangle}}\right)^{1/2}}{\left\langle|\bm{B} - \langle\bm{B}\rangle|^2\right\rangle}},
\end{equation}
which measures how the components of $\bm B$ are correlated to keep $|\bm B|$ constant. $C_B^2=0$ for perfectly spherically polarized fluctuations. 
The initial condition is chosen to mimic a strongly $z^+$-dominated, nearly spherically polarized Alfvénic wind propagating outwards from near the Alfvén point: $\bm z^+(t=0)$ carries the fluctuation energy while $\bm z^-(t=0)=0$. Radial expansion and large-scale inhomogeneity produce partial reflection (creation of $\bm z^-$) as the packet propagates outward, thereby initiating reflection-driven turbulence.   We initialize the fluctuations as linearly polarized Alfvén waves ($\delta \bm{u}=\delta \bm{B}/\sqrt{4 \pi \rho}$, corresponding to $\bm{z}^-=0$) with $\delta \bm u$ and $\delta \bm B$ lying in the $(\bm k\times {{\bar{\bm B}}})$ direction \citep{Squire2020,Johnston2022}. 
The initial set of Alfvén waves are generated from a sum of randomly phased waves with amplitude determined by Gaussian energy spectrum,
\begin{equation}
    E(k_\parallel,k_\perp) \propto \exp\!\Bigl[-\frac{(k_\parallel - k_{\parallel,0})^2 + (k_\perp - k_{\perp,0})^2}{k_w^2}\Bigr].  
\end{equation}
Here $k_{\parallel,0}=\kappa_\parallel({2\pi}/{L_x}),\; k_{\perp,0}=\kappa_\perp({2\pi}/{L_\perp}),\; \text{and}\;  k_w=({12}/{L_\perp})$, which set the scale of the center of Gaussian peak and the width of the peak respectively.
$k_{\parallel,0}$ and $k_{\perp,0}$ lie respectively parallel and perpendicular to $\bar{\bm B}$, maintaining anisotropy even with an initial tilted spiral field. As the expanding-box simulation begins, the modes rapidly adjust within approximately 0.5 Alfv\'en times, losing the variation in $|\bm B|$ caused by their initial linear polarization. This natural relaxation reduces the magnetic compressibility $C_B^2$ and allows the fluctuations to become spherically polarized.  

To choose simulation parameters and quantify regimes, we use the timescales and their ratios, as discussed in \cref{Theory}. The expansion time is $\tau_{\rm exp}^{-1} = \dot a / a$, the linear timescale is $\tau_{\rm A}^{-1} = k_\parallel v_{\rm A}$, and the outer-scale nonlinear time is $\tau_{\rm nl}^{-1} = k_{\perp} z^+$. From these rates we form the dimensionless control parameters: (i) $\chi_{\rm exp,0} = (k_{\perp,0} \; z^+_{\rm rms0}) / (\dot a/a)$; (ii) $\chi_{\rm A} \doteq (k_\perp z^+)/(k_\parallel v_{\rm A})$, which compares strength of nonlinear interactions to linear Alfvénic propagation and thus quantifies how strongly nonlinearity competes with Alfvénic de-correlation of $z^-$ fluctuations (\citealt{GS95}); and (iii) the ratio of box-scale Alfvén time to the expansion time $\Delta_0 \doteq \chi_{\rm exp0}/\chi_{\rm A0} = v_{\rm A,0}/\dot{a} L_x$ which is constant. For the parameter set used in this paper we fix $\Delta_0 = 2$ (via $\dot{a}=0.5$, $L_\parallel=1$ and $v_{\rm A}=1$) and vary $\chi_{\rm exp,0}$ roughly in the range $\approx 11\!-\!30$. 
We explore two initial fluctuation amplitudes defined by \(A\equiv { z^+_{\rm rms0}}/{v_{\rm A,0}}=1.0 \;\text{and}\; 0.5,\) where $z^+_{\rm rms0}$ is the initial root-mean-square fluctuation amplitude (see Table~\ref{tab1:sim_params} for exact values).   The moderate-resolution simulations with reduced amplitude ($A_0 = 0.5$) are designated as A05 to distinguish them from the standard moderate-resolution runs with $A_0 = 1.0$.   Changing \( A\) alters the initial nonlinear time (and therefore \(\chi_{\rm exp,0}\) and \(\chi_{\rm A,0}\)), so we fix \(\chi_{\rm exp,0}\) to isolate pure amplitude effects. Physically, larger $A$ produces larger fluctuations with larger $\delta B_\parallel$ to maintain their spherical polarization, whereas smaller $A$ (initial conditions) remain closer to the reduced MHD limit, which is derived assuming $z^+/v_{\rm A}\ll1$.

\begin{table}
  \centering
  \caption{Simulation parameters for the expanding‑box runs. High-resolution (HR) runs use $1200^3$ grid points; moderate-resolution (MR) runs use $720^3$. The parameters explored are; \(\Phi_0\) is the initial PS angle, \(L_{x0}/L_{\perp0}\) the initial box aspect ratio (where $L_{\perp0}$ denote both $L_{y0}$ and $L_{z0}$), \(\dot a\) the expansion rate, \(A_0 \) the initial fluctuation amplitude, and \(\beta_0\) the plasma beta. ${k}_{\rm peak}$ set the center of Gaussian peak in $k$-space, \(\chi_{\rm exp,0}\) the initial expansion-to-nonlinearity ratio, and \(k_{\rm w}\) set the width of the Gaussian peak.}
  \label{tab1:sim_params}
  \begin{tabular}{@{}lccccccccc@{}}
    \toprule
    \textbf{Simulation} & \textbf{Resolution} & $\Phi_0 $ & $L_{x0}/L_{\perp 0}$ & $\dot{a}$ & $ A_0$ & $\beta_0$ & $\kappa_{\rm peak}$ & $\chi_{\rm exp,0}$ & $k_{\rm w}$ \\
    \midrule
    HR-$\Phi_0\!=\!0^\circ$     & $1200^3$ & $0^\circ$     & $5$ & $0.5$ & $1.0$ & $0.3$ & $(1.5, 1.5)$ & $\approx 15$ & $12.0$ \\
    HR-$\Phi_0\!=\!4^\circ$     & $1200^3$ & $4^\circ$     & $5$ & $0.5$ & $1.0$ & $0.3$ & $(1.5, 1.5)$ & $\approx 15$ & $12.0$\\
    
    A05-$\Phi_0\!=\!0^\circ$    & $720^3$  & $0^\circ$     & $5$ & $0.5$ & $0.5$ & $0.3$ & $(3.0, 3.0)$ & $\approx 15$ & $12.0$\\
    A05-$\Phi_0\!=\!1.5^\circ$  & $720^3$  & $1.5^\circ$   & $5$ & $0.5$ & $0.5$ & $0.3$ & $(3.0, 3.0)$&  $\approx 15$ & $12.0$\\
    A05-$\Phi_0\!=\!2^\circ$  & $720^3$  & $2^\circ$   & $5$ & $0.5$ & $0.5$ & $0.3$ & $(3.0, 3.0)$&  $\approx 15$ & $12.0$\\
    A05-$\Phi_0\!=\!5^\circ$   & $720^3$  & $5^\circ$    & $5$ & $0.5$ & $0.5$ & $0.3$ & $(3.0, 3.0)$ & $\approx 15$ & $12.0$\\

    MR-$\Phi_0\!=\!0^\circ$  & $720^3$ & $0^\circ$   & $5$ & $0.5$ & $1.0$ & $0.3$ & $(2.0, 2.0)$ & $\approx 45$ & $60.0$\\
    MR-$\Phi_0\!=\!2^\circ$  & $720^3$ & $2^\circ$   & $5$ & $0.5$ & $1.0$ & $0.3$ & $(2.0, 2.0)$ & $\approx 45$ & $60.0$\\
    MR-$\Phi_0\!=\!5^\circ$  & $720^3$ & $5^\circ$   & $5$ & $0.5$ & $1.0$ & $0.3$ & $(2.0, 2.0)$ & $\approx 45$ & $60.0$\\
    \bottomrule
   \end{tabular}
  
\end{table}
\subsection{Numerical issues at large $a$}\label{Numerical issues}
We note that a numerical instability appears in our PS runs once the local spiral angle becomes very large ($\gtrsim 70^\circ$), which occurs at sufficiently large $a$. The instability manifests as small-scale, speckle-like noise in the transverse fields and an abrupt spike in inward Elsässer energy. The underlying cause is uncertain, but a plausible explanation is the increasing obliquity of the mean field along with the extreme anisotropic deformation of the EBM domain (the initially elongated box becoming a very flattened pancake), which under-resolves transverse modes and amplifies numerical error. This unfortunately stops us accessing any possible final balanced phase in PS runs. We therefore exclude affected intervals from our analysis. The issue is discussed in more detail in Appendix~\ref{appendixB}.

\section{Results}\label{theoretical validation}
\subsection{Turbulence Evolution and Emergence of the Parker Spiral}
\begin{figure}
        \centering
        \includegraphics[width=\linewidth]{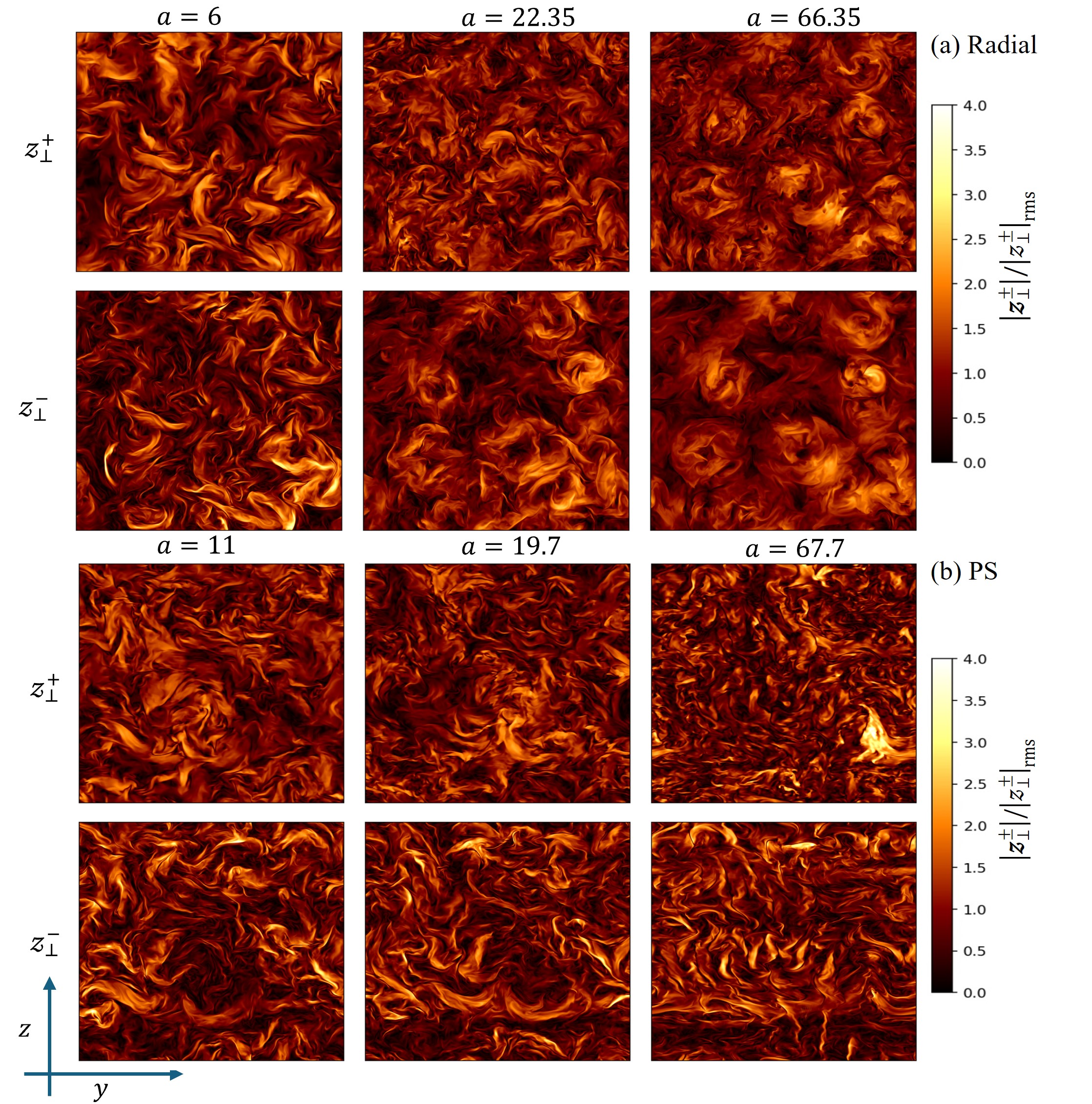}
\caption{Snapshots of the Elsässer fields $|\bm z_\perp^\pm|$ perpendicular to magnetic field in the $y$-$z$ plane. The top two rows (a) show different stages of expansion are shown for the A05-$\Phi_0\!=\!0^\circ$ with a radial $\overline{\bm B}$ simulation at $a \approx 6$ (left), $a \approx 22.35$ (middle), and $a \approx 50.35$ (right). These snapshots illustrate the turbulent evolution from an initially imbalanced regime to a magnetically dominated and balanced phase.
The bottom two rows (b) show the snapshots of the Elsässer fields in the PS case for A05-$\Phi_0\!=\!1.5^\circ$.
Elsässer fields $|\bm z_\perp^\pm|$ are shown at three stages of expansion: $a \approx 11$ (left), $a \approx 19.7$ (middle), and $a \approx 67.7$ (right) with the spiral angles of $\Phi \approx 16.1^\circ, \; 27.3^\circ, \; 60.57^\circ$ respectively. The system remains turbulent, with distinctive anisotropic structural features due to the component of the mean magnetic field along $y$. Note that in each panel, fluctuations are normalized by their rms value of each time.}
\label{fig:snaps720R}
\end{figure}

\begin{figure}
    \centering
    \begin{subfigure}[b]{\linewidth}
        \includegraphics[width=\linewidth]{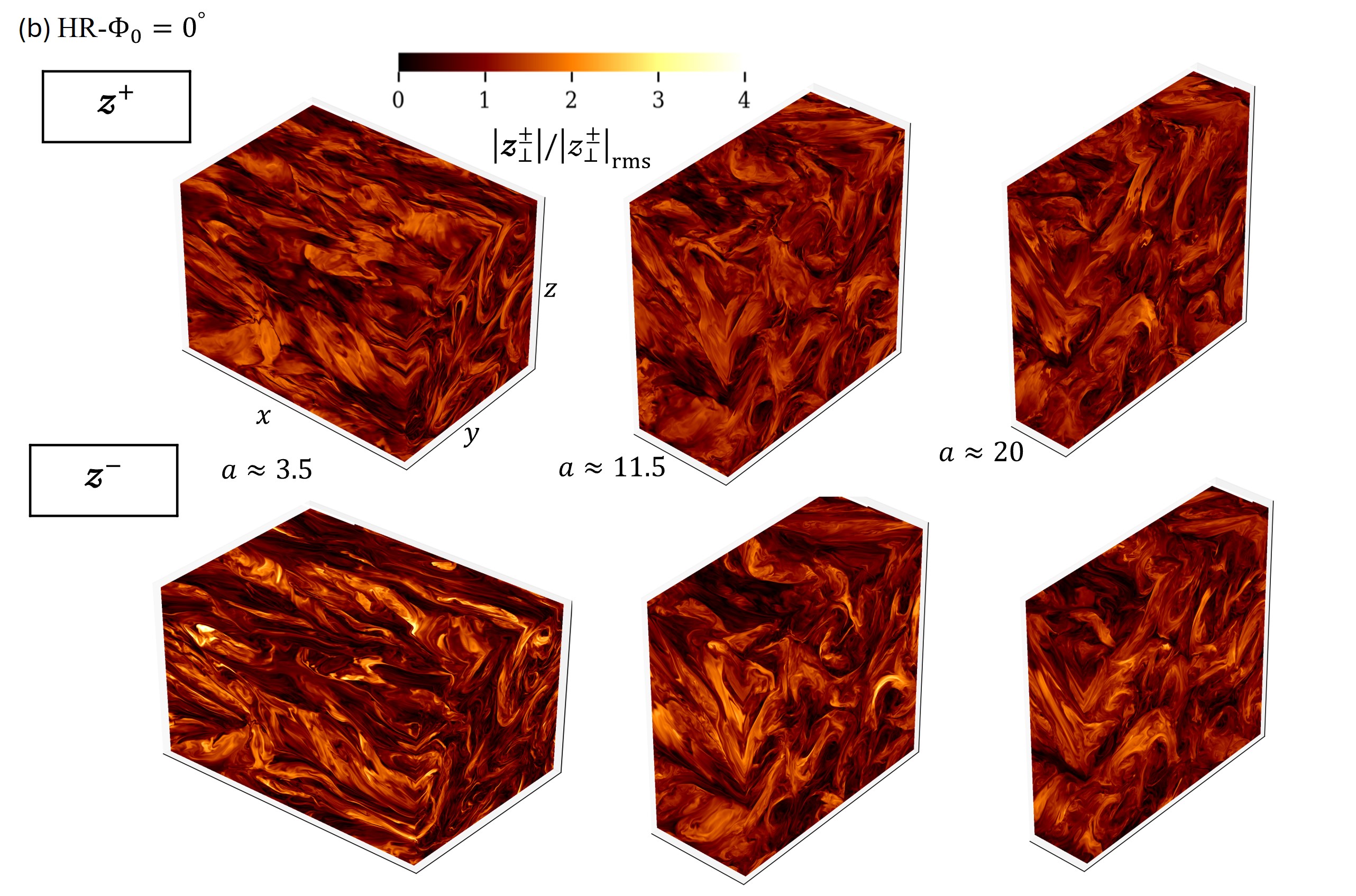}
    \end{subfigure}
    \hfill
        \begin{subfigure}[b]{\linewidth}
        \includegraphics[width=\linewidth]{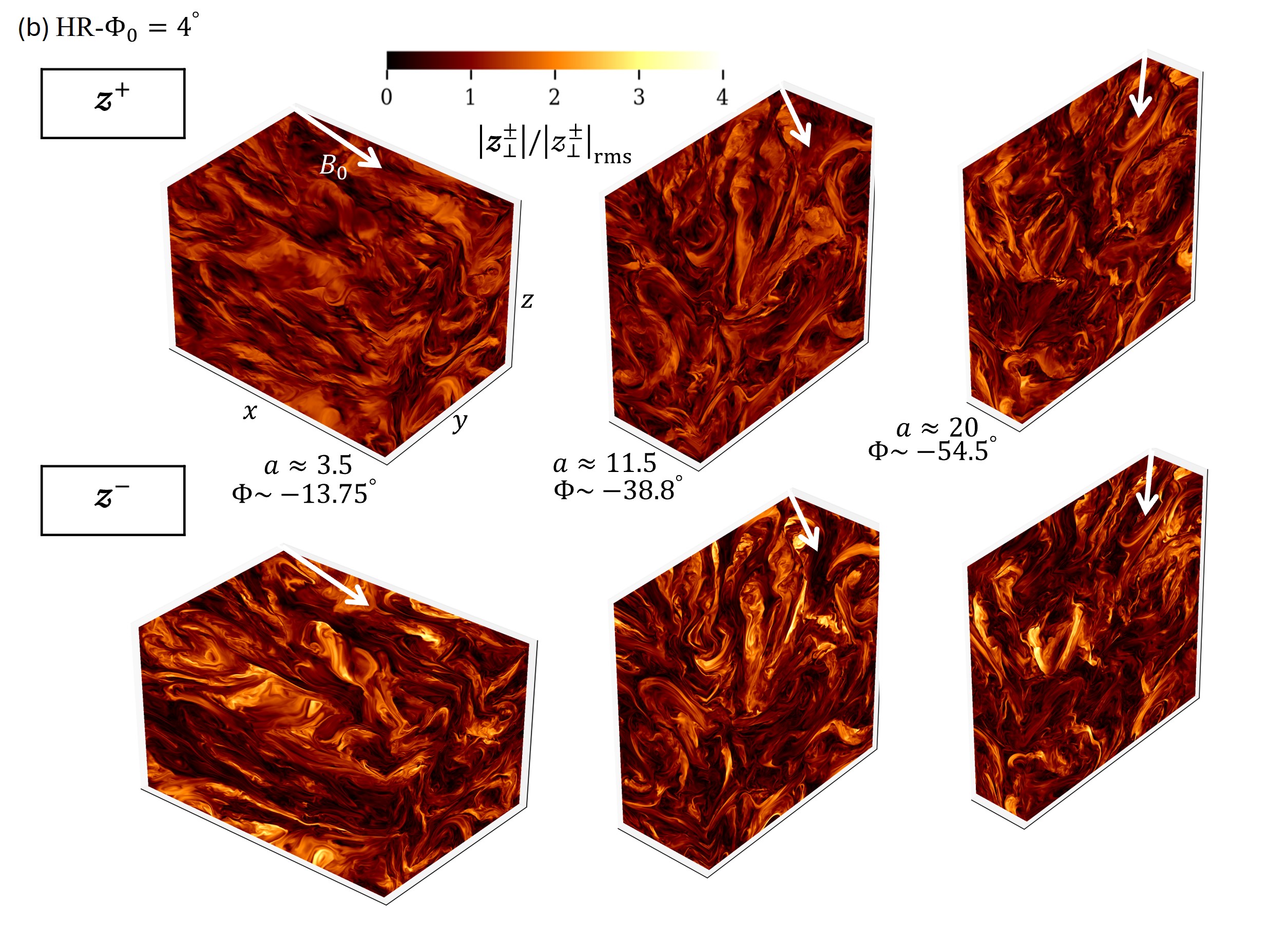}
    \end{subfigure}
    \caption{3D visualizations of the Elsässer fields $|\bm z_\perp^\pm|$ perpendicular to the magnetic field at $a \approx3.5, \; 11.5,\; 21$ for HR-$\Phi_0\!=\!0^\circ$ and HR-$\Phi_0\!=\!4^\circ$ simulations. White arrow shows the direction of mean magnetic field in PS run.} 
    \label{fig:z_perp_snapshots}
\end{figure}

The morphology of turbulence depends sensitively on the background magnetic geometry. The top panels of Fig.~\ref{fig:snaps720R}a depict two-dimensional snapshots ($y$-$z$ mid plane slice) of the perpendicular normalized Elsässer fields $|\bm z_\perp^\pm|$ for the radial A05-$\Phi_{0}\!=\!0^\circ$ run (here $\bm z_\perp^\pm = (\bm I - \hat{\bm b}\hat{\bm b})\cdot\bm z^\pm$, which is approximately the Alfvénic part). In the radial case, turbulence evolves in two distinct phases. During the early, imbalanced stage when $\chi_{\rm exp}>1$, outward $z^+$-fluctuations are strong and develop fine perpendicular structure, while reflected $z^-$ fluctuations appear as more intermittent, field-aligned filaments. As expansion proceeds and $\chi_{\rm exp}$ approaches unity, reflected normalized wave-action energy grows and $\tilde E^-/\tilde E^+$ approaches unity, with both fields acquiring increasingly similar circular vortex-like structures. In this balanced phase, the nonlinear transfer rate decreases relative to expansion and the cascade stalls, leaving coherent vortices to dominate the dynamics and shutting off heating. Throughout both stages, expansion stretches eddies across the mean field, generating large-aspect-ratio quasi 2D eddies towards the simulation's end. This behavior qualitatively matches the reduced-MHD results of \citet{Meyrand2025}, who reported 2D ``Alfvén vortices'', similar to those we see, in the late stage $\chi_{\rm exp} \sim 1$ magnetically dominated regime.

The Parker spiral simulations evolve differently. In A05-$\Phi_{0}\!=\!1.5^\circ$ (lower two panels of Fig.~\ref{fig:snaps720R}), turbulence does not relax toward a quasi-two-dimensional state as in the radial case. Instead, nonlinear activity persists to larger expansion factors because (as we will show below) $\chi_{\rm exp}$ remains larger. The fields develop persistent, ribbon-like structures with sharper transverse gradients and smaller $\ell_\perp$ than in radial runs (this will be quantified in \cref{Scales}). Physically, this follows because the continuous rotation of the mean field during expansion maintains a persistent misalignment between the perpendicular and expansion directions; as illustrated in Fig.~\ref{fig:EMB_spiral_expansion} and discussed in \cref{sub:PS Theory}, that misalignment prevents eddies from maintaining cylindrical symmetry about the local field and thereby inhibits the formation of quasi-2D vortices. Consequently, we observe a persistent, irregular 3D anisotropy in PS runs and persistent turbulent heating. 

Figure~\ref{fig:z_perp_snapshots} shows similar features via a three-dimensional depiction of the Elsässer fields from the high-resolution (HR-$\Phi_0\!=\!0^\circ$ and HR-$\Phi_0\!=\!4^\circ$) simulations. These volumetric views support the conclusion of Fig.~\ref{fig:snaps720R}, and also show the evolution of radial structure. In HR-$\Phi_0\!=\!0^\circ$, turbulence condenses into slowly developing Alfvén vortices, though these are not so obvious here because of the earlier time. In HR-$\Phi_0\!=\!4^\circ$, structures stay at smaller scale and are possibly more intermittent, developing 3D outer-scale anisotropy. The resulting sustained small $\ell_\perp$ inhibits the cascade from stalling. We also see qualitative agreement with prior studies linking the PS geometry to 3D anisotropy \citep{Verdini2019}.

\subsection{Energy Evolution and Heating}
\begin{figure}
\centering
        \includegraphics[width=0.9\linewidth]{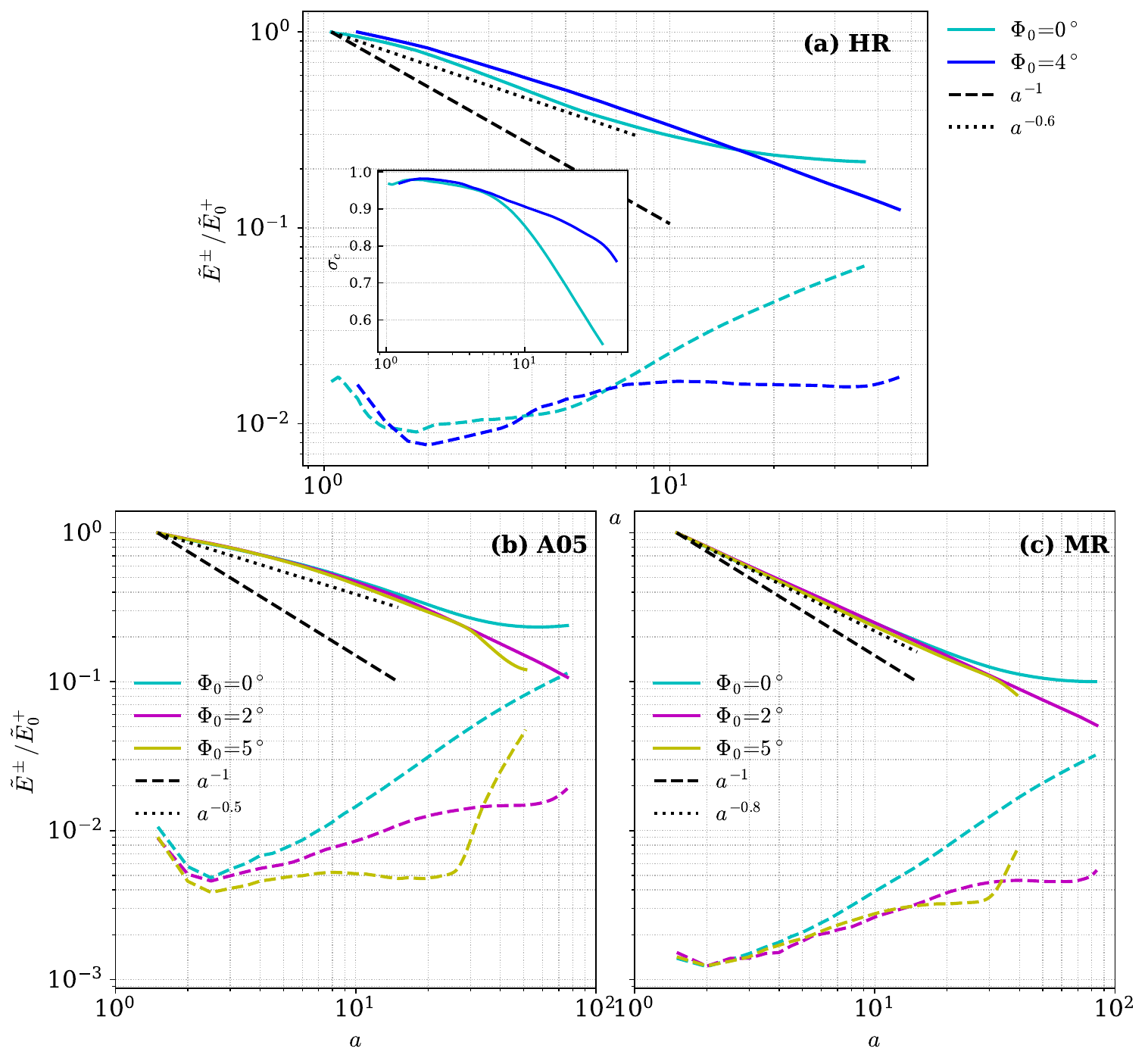}
    \caption{ Evolution of normalized Elsässer wave action energies, $\tilde E^\pm(a)/\tilde E^+_0$, as a function of $a$ . Solid curves denote outward wave-action energy ($\tilde{E}^+$) and dashed curves denote inward wave-action energy ($\tilde{E}^-$); line colors correspond to different initial PS angles. Panel (a) shows the high–resolution runs (HR-$\Phi_0\!=\!0^\circ$ and HR-$\Phi_0\!=\!4^\circ$). In the radial case ($\Phi_0\!=\!0^\circ$), the outward energy decays approximately as $a^{-0.6}$ up to $a\sim15$, after which $\tilde E^+$ flattens and rises slightly, indicating that turbulent heating effectively shuts off. In the PS runs, the outward energy continues to decay to larger $a$, maintaining an imbalanced cascade and sustained dissipation for longer. The inset displays the normalized cross helicity $\sigma_c$ for these runs. Panel (b) shows A05-$\Phi_0\!=\!0^\circ,2^\circ,5^\circ$ runs with similar initial $\chi_{\rm exp,0}$ as panel (a) but reduced amplitude $\rm A=0.5$. The lower amplitude, near-RMHD case decays slightly more slowly than the high amplitude case, roughly as $\tilde E^+\propto a^{-0.5}$, possibly because the higher  amplitude spherical-polarized fluctuations help to make reflection more efficient. (c) MR-$\Phi_0\!=\!0^\circ,2^\circ,5^\circ $ runs with higher initial $\chi_{\rm exp,0}$, which leads to steeper decay following approximately $\tilde E^+\propto a^{-0.8}$. The runs with larger spiral angles (e.g., $\Phi_0\!=\!5^\circ$) were terminated at smaller distances due to numerical instabilities discussed in \cref{Numerical issues}. Note that the first snapshot $(a=1)$ is omitted to improve visualization, since the inward-propagating mode $\tilde z^-$ is negligible at that stage.}
\label{fig:energy_evolution}
\end{figure}
Figure~\ref{fig:energy_evolution} shows the normalized wave action energies, $\tilde E^\pm/\tilde E^+_0$, as functions of expansion factor $a$. Here $\tilde{E}^+_0 = \tilde{E}^+(a=1)$ is the initial outward wave-action energy, so its decrease provides a measure of turbulent heating per \eqref{heating_rate}. Plotted are HR-$\Phi_0\!=\!0^\circ,4^\circ$; A05-$\Phi_0\!=\!0^\circ,2^\circ, 5^\circ$, and MR-$\Phi_0\!=\!0^\circ,2^\circ,5^\circ$ runs spanning initial $\chi_{\rm exp,0}$ in $[\approx 15,\; 45]$. 
For runs with smaller \(\chi_{\rm exp,0}\) we find an initial approximate power-law decay
\(\tilde E^+\propto a^{-0.6}\) or $a^{-0.5}$; for larger
\(\chi_{\rm exp,0}\) the decay is faster with \(\sim a^{-0.8}\). This discrepancy with the \(a^{-1}\) limit predicted in RDT phenomenology of \cref{Phenomnology} (Eq.~\ref{energy_prediction}) is likely because the $a^{-1}$ decay is recovered only for sufficiently large \(\chi_{\rm exp,0}\). A similar conclusion was found in \citet{Meyrand2025}, who found that very large $\chi_{\rm exp}$ was needed to recover the expected $a^{-1}$ decay.

Focusing on the HR runs, the radial case displays a slow change near $a\!\sim\!15$ where $\tilde E^+$ stops decaying. We will show below that this is consistent with $\chi_{\rm exp}$ approaching unity, as discussed in the section \ref{Theory}. Concurrently $\tilde E^-$ grows with expansion, following a similar trend, and approaching $\tilde E^+$ as the system moves towards a quasi-balanced state. During this transition from a strongly imbalanced turbulent state to a quasi balanced sate, the nonlinear coupling weakens because $\bm z^+$ and $\bm z^-$ becomes proportional (see Fig.~\ref{fig:snaps720R} and \ref{fig:z_perp_snapshots}) and the turbulent heating rate drops. Eventually, as shown by \citet{Meyrand2025} we expect $\tilde E^+$ to rise as the system starts developing a very large negative residual energy,
\begin{equation}
\label{eq:sigma_r}
\sigma_r \;=\; \frac{\langle \lvert\delta\bm{u}\rvert^2 \rangle - \langle \lvert\delta\bm{b}\rvert^2 \rangle}
{\langle \lvert\delta\bm{u}\rvert^2 \rangle + \langle \lvert\delta\bm{b}\rvert^2 \rangle}
\;=\; \frac{\langle \bm{z}^-\!\cdot\!\bm{z}^+ \rangle}{\langle \lvert\bm{z}^+\rvert^2 \rangle + \langle \lvert\bm{z}^-\rvert^2 \rangle},
\end{equation}
which causes wave-action energy to grow as $\tilde{E}^\pm\propto a$ (see Eq.~\eqref{energy_eq}).

In the PS run, the outward energy decays beyond $a\!\sim\!15$ without an observable slowing of its decay. This power-law decay in $\tilde E^+$ thereby sustains nonlinear turbulent heating to larger heliocentric distances than in the radial case. Below, we show that this is due to a geometry-driven modification of perpendicular scales: the PS runs exhibit a slower growth (or saturation) of $\ell_\perp$, which decreases $\tau_{\rm nl}$ relative to expansion and keeps $\chi_{\rm exp} >1$ rather than it collapsing to unity. As a result, turbulent dissipation remains active, and the cascade persists over significantly larger heliocentric distances. Likewise, as expected from the RDT prediction $\tilde z^-/\tilde z^+ \sim \chi_{\rm exp}^{-1}$ (see Eq.~\ref{RDT Prediction}), $\tilde E^-$ remains smaller than in the radial case, the normalized cross-helicity, 
\begin{equation}
\label{eq:sigma_c}
\sigma_c \;=\; \frac{\langle \lvert\bm{z}^+\rvert^2 \rangle - \langle \lvert\bm{z}^-\rvert^2 \rangle}
{\langle \lvert\bm{z}^+\rvert^2 \rangle + \langle \lvert\bm{z}^-\rvert^2 \rangle},
\end{equation}
stays higher (inset of Fig.~\ref{fig:energy_evolution}a), and the system retains a strongly Alfvénic, imbalanced state over larger distances.

The simulations also allow a cursory examination of the influence of fluctuation amplitude on the energy evolution. Figure~\ref{fig:energy_evolution} compares (a) a high-amplitude case with $\rm A=1.0$ and (b) a lower amplitude case with $\rm A=0.5$ and similar $\chi_{\rm exp,0}$. In the lower amplitude case, the fluctuations remain nearly linearly polarized (reduced-MHD like), and follow a slower energy decay $\tilde E^+ \propto a^{-0.5}$, while higher-amplitude runs, which must have larger $\delta B_\parallel$ to maintain spherical polarization exhibit steeper decay. 
One possible mechanism to explain this difference is that, the anisotropic reflection term (see Eq.~\ref{eq:wave action}) that seeds \(\tilde{\bm z}^-\) from \(\tilde{\bm z}^+\) modifies the character of the reflection once fluctuations become large. For small-amplitude, nearly transverse Alfvénic fluctuations, the reflection operator \(\tilde{\mathbb{T}}=\mathrm{diag}(-1,1,1)\) leads to \(\tilde z^-_{y,z}\propto -\tilde z^+_{y,z}\) with \(\tilde z^\pm_x\) negligible. At large amplitude, however, finite \(\delta B_\parallel\) (nonzero \(\tilde z_x^\pm\)) and the oblique mean field mix components: the \(-1\) in \(\tilde{\mathbb{T}}\) produces additional contributions to \(\tilde z_x^- \propto \tilde z_x^+\). In other words, high amplitude fluctuations scramble the simple \(\tilde{\bm z}^-\propto -\tilde{\bm z}^+\) relation, reduce the magnitude of \(\langle\tilde{\bm z}^+\!\cdot\!\tilde{\bm z}^-\rangle\), and thereby allow stronger nonlinearity to persist in the PS geometry compared with the radial case. In-situ near-Sun measurements often show large normalized fluctuation amplitudes (\citealt{Bale2019}), the high-amplitude runs are likely more representative of observed solar wind.

\subsection{Evolution of length-scales and $\chi_{exp}$}\label{Scales}
The ratio of expansion to nonlinear timescales, $\chi_{\rm exp}$ governs whether nonlinear interactions remain strong enough to sustain a cascade. Because $\tau_{\rm nl}\sim \ell_\perp/z^\mp$, the evolution of $\chi_{\rm exp}$ is directly tied to the growth of perpendicular correlation lengths, and the decay of amplitude $z^+$.

To compute directional correlation lengths we follow the procedure below. First, we draw an ensemble of straight periodic lines in the direction of $\hat{\ell}\in{\{\hat{e}_\parallel,\hat{e}_T,\hat{e}_N}\}$ through the computational domain and evaluate the interpolated magnetic fluctuation \(\delta\bm{B}(s)\) along each line. To isolate the Alfvénic contribution we form the projection
\begin{equation}
    \delta \bm B_{\rm A}(s)=(\hat{b}\times\hat{\ell})\cdot\delta\bm{B}(s),
\end{equation} 
Each retained line \(f_n=\delta B_{\rm A}(s_n)\) is demeaned and the autocorrelation $C_f[k]$ is computed, then normalized by $C_f(0)$ to yield the per-line correlation; these per-line correlations are then ensemble-averaged to produce the final correlation function $\mathcal{C}(s)$.
The characteristic correlation lengths \(\ell\) are then extracted from $\mathcal{C}(s)$ using the integral estimator \(\ell_{\rm int}=\int_0^{s_{\rm cut}}\mathcal{C}(s) ds\) with \(s_{\rm cut}\) taken at the first zero crossing. These \(\ell\) values are reported for the Alfvénic projection \(\delta B_{\rm A}\). 

\begin{figure}
    \centering
    \includegraphics[width=\linewidth]{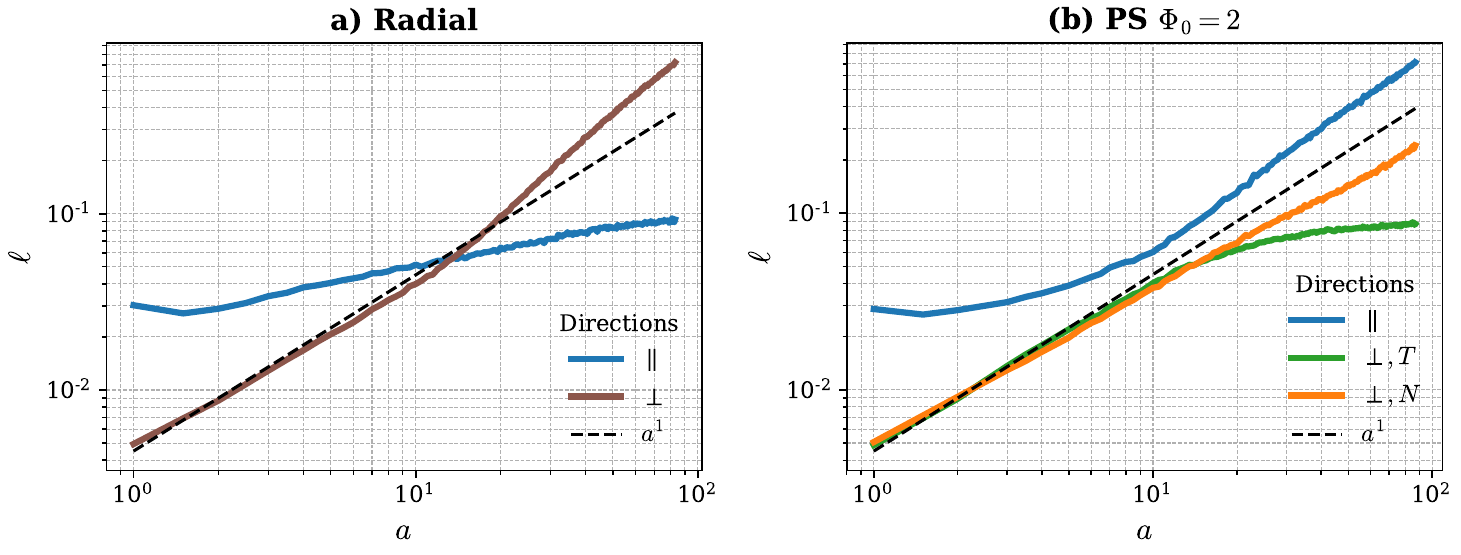}
    \caption{Evolution of correlation length $\ell$ as a function of $a$ for magnetic-field ($\bm{B_\perp}$) fluctuations for MR-$\Phi_0\!=\!0^\circ$ and MR-$\Phi_0\!=\!2^\circ$ runs. Each curve represents projections along the field ($\parallel$, blue), perpendicular($\perp$, brown), $\hat{e}_T$ direction ($\perp,T$, green), and $\hat{e}_N$ direction ($\perp,N$, orange). The dashed black line indicates the $a^{1}$ power-law scaling for reference. In the radial case (left panel), $\ell_{\perp}$ grows nearly linearly with $a$, consistent with expansion-driven eddy widening at early times then transitions to faster growth at late times (see \citealt{Meyrand2025}). In contrast, the Parker spiral geometry modifies the growth: $\ell_{\perp,T}$ saturates and remains nearly constant, while $\ell_{\perp,N}$ increases more slowly than linear.}
    \label{fig:corr_lengths}
\end{figure}
Figure~\ref{fig:corr_lengths} shows the correlation lengths of magnetic fluctuations in both radial and PS runs.
In the radial case, the perpendicular correlation length ($\ell_{\perp}$), which is the average of those $\hat{e}_T$ and $\hat{e}_N$, initially grows roughly proportional to $a$ as expected for uniform expansion. At later times, however, as the flow becomes magnetically dominated and quasi-two-dimensional, the co-moving $\ell_{\perp}$ begins to grow even faster than this linear $a$ scaling (Fig.~\ref{fig:corr_lengths}a). This accelerated broadening of structures produces flattened, pancake-like eddies, steadily reducing $k_\perp$ and increasing the nonlinear time $\tau_{\rm nl}$. Visually this phase corresponds to the dominance of coherent Alfvénic vortices, growing in co-moving transverse size as the $z^+$ amplitude decreases. This behavior aligns with the phenomenology proposed by \cite{Meyrand2025} (Fig. 5), who argued that conservation of wave-action anastrophy---the squared magnetic vector potential in wave-action variables---drives a split transfer of energy to larger perpendicular scales. In our compressible simulations, the same qualitative behavior emerges, consistent with Alfvén vortices conserving anastrophy on intermediate timescales and expanding even faster than expansion as they decay.

In contrast, Parker spiral geometry alters this scaling at larger $a$, as expected from theory \cref{sub:PS Theory} and Fig.~\ref{fig:Theory}. The tangential correlation length saturates, while the normal length grows more slowly than linearly, signaling the reduction of eddy scale in the co-moving frame. This anisotropy prevents indefinite transverse stretching, limits the increase of $\tau_{\rm nl}$, and thereby sustains turbulent activity. 
 \begin{figure}
  \centering
  \includegraphics[width=\linewidth]{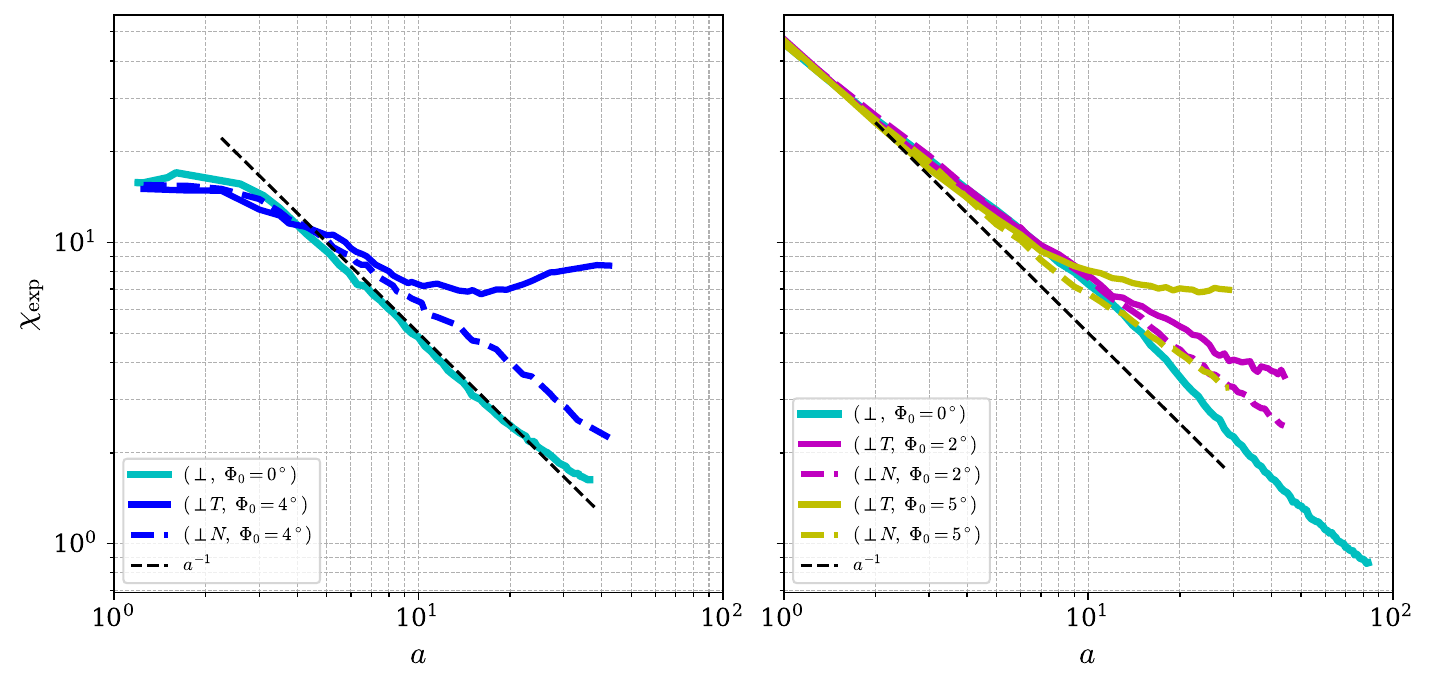}
  \caption{The time-scale ratio $\chi_{\rm exp}$ plotted as a function of expansion factor $a$. We computed this using Alfvénic perpendicular correlation lengths for magnetic field fluctuations for HR-$\Phi_0 = 0^\circ,4^\circ$ (left) and MR-$\Phi_0= 0^\circ, 2^\circ,5^\circ$ simulations (right). The solid lines show $\chi_{\rm exp}$ calculated from $\ell_{\perp ,T}$, and the dotted lines show $\chi_{\rm exp}$ computed from $\ell_{\perp ,N}$. In the radial case, we averaged the both to compute $\ell_{\perp }$, and  $\chi_{\rm exp}$ decays as $a^{-1}$. In contrast, the PS geometry breaks this symmetry: both $\ell_{\perp,T}$ and $\ell_{\perp,N}$ remain smaller than in radial case yielding systematically larger $\chi_{\rm exp}$. This sustains turbulence at larger $a$ because $\chi_{\rm exp}$ remains greater than unity for longer range.}
 \label{fig:chi_exp_evolution}
\end{figure}

The consequences for $\chi_{\rm exp}$ are shown in Fig.~\ref{fig:chi_exp_evolution} for the HR-$\Phi_0\!=\!0^\circ,4^\circ$ (left) and MR-$\Phi_0\!=\!0^\circ,2^\circ,5^\circ$ simulations (right). All runs show an initial decay of $\chi_{\rm exp}$ as amplitudes drop and scales expand. In the radial case, $\chi_{\rm exp}$ decreases nearly as $a^{-1}$ as expected from theory (see \cref{Phenomnology}). The somewhat larger $\chi_{\rm exp}$ at $a\simeq 20$ in MR compared to HR runs is consistent with the delayed flattening of the decay of $\tilde{E}^+$ in the former, although it would be very valuable to explore a wider range in $\chi_{\rm exp0}$ for future work, to study this in more detail.

The Parker spiral cases diverge from this trend at larger $a$. Because the tangential scale stops growing, $\tau_{\rm nl}$ does not diverge, and $\chi_{\rm exp}$ starts decaying more slowly or even growing, when it is still above unity. This change in geometry delays the approach to balance and preserves the cascade over a much broader radial extent, with more sustained plasma heating relative to the radial case. The late-time separation between tangential and normal scales is a another interesting feature of the PS-induced anisotropy. We note, however, that our PS simulations could not be extended to later times due to numerical limitations, which prevent us from fully capturing the asymptotic evolution of the cascade. 

\begin{figure}
    \centering
    \includegraphics[width=\linewidth]{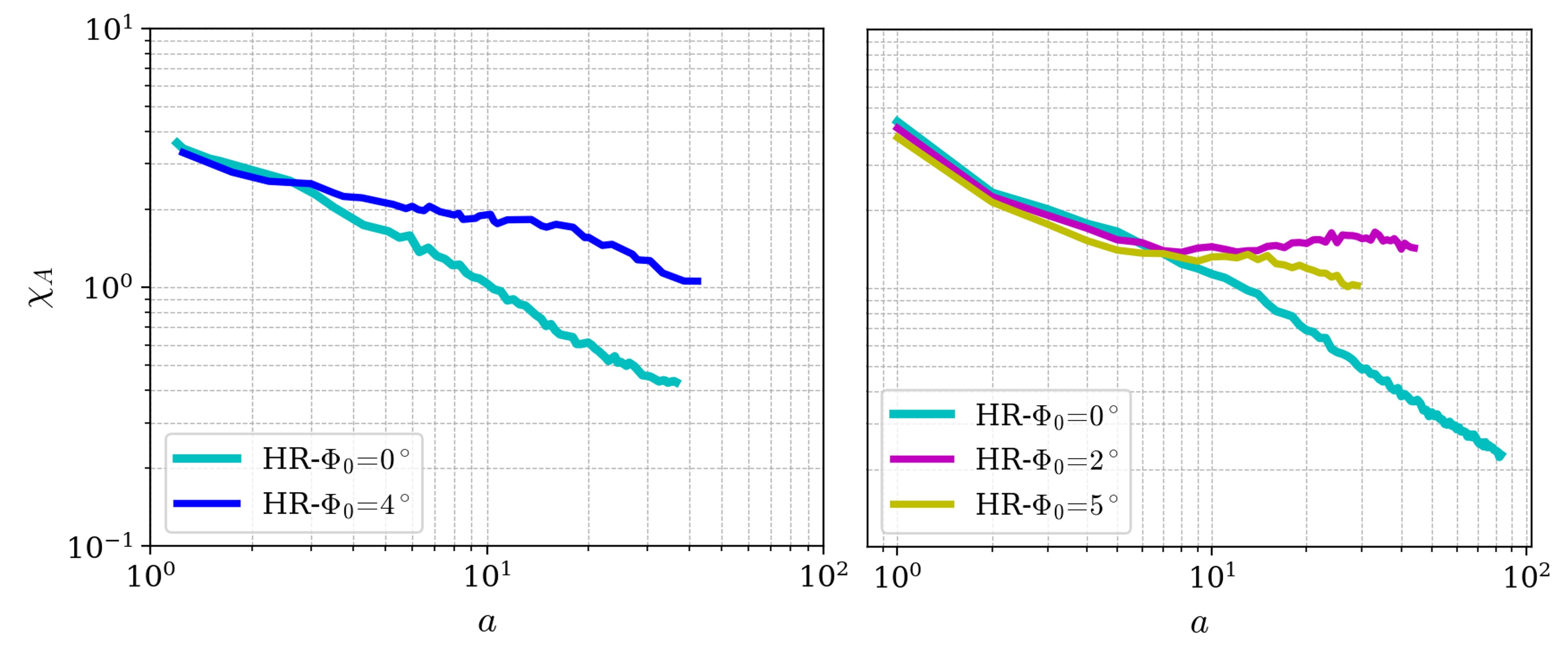}
    \caption{Measured $\chi_{\rm A}$  versus $a$. Note that, we take the average of transverse and normal perpendicular correlation lengths $\ell_\perp = (\ell_{\perp,T} + \ell_{\perp,N})$ for this figure.}
    \label{fig:chi_A}
\end{figure}
We also measure the critical balance parameter of the turbulent cascade using $\chi_{\rm A} \simeq (k_\perp  z^+)/(k_\parallel v_{\rm A}) = (\ell_\parallel z^+)/(\ell_\perp v_{\rm A})$. An assumption of the RDT phenomenology (discussed in \cref{Phenomnology})---that the deriving due reflection source $\mathcal{R}^- \tilde z^+$ balances the nonlinear sink ($\mathcal{N}^- \tilde{z}^-$) in Eq.~\eqref{eq:cartoon}---likely requires $\chi_{\rm A}\gtrsim 1$. Only for $\chi_{\rm A}$ of order unity or larger do we expect nonlinear interactions to dominate, allowing a strong turbulent cascade to develop and damp fluctuations as assumed. In the opposite limit of $\chi_{\rm A}\gtrsim 1$ fluctuations are expected to rapidly de-correlate in $k_\parallel$ and drop to $\chi_{\rm A}\sim1$ \citep{Schekochihin2022}. In Fig.~\ref{fig:chi_A}, we thus expect the parallel scales to adjust to cause turbulence to sit at $\chi_{\rm A}\sim 1$. While there is only tentative evidence for this in the radial runs---we see a drop to $\chi_{\rm A}\approx 1.5\rightarrow2$ initially but a larger $\chi_{\rm exp0}$ would be needed to allow a longer decay phase where this could be studied in detail---the PS runs sit near $\chi_{\rm A} \sim 1$ at late times, consistent with sustained, critically balanced turbulence.   The radial runs trend toward a true 2D state with $k_\parallel=0$ ($\ell_\parallel \rightarrow \infty$) at late times, but the measured $\ell_\parallel$ becomes box-limited (2D modes are not well captured by method)   and the plotted $\chi_{\rm A}$ is therefore not necessarily a useful measure.

\begin{figure}
    \centering
    \includegraphics[width=0.75\linewidth]{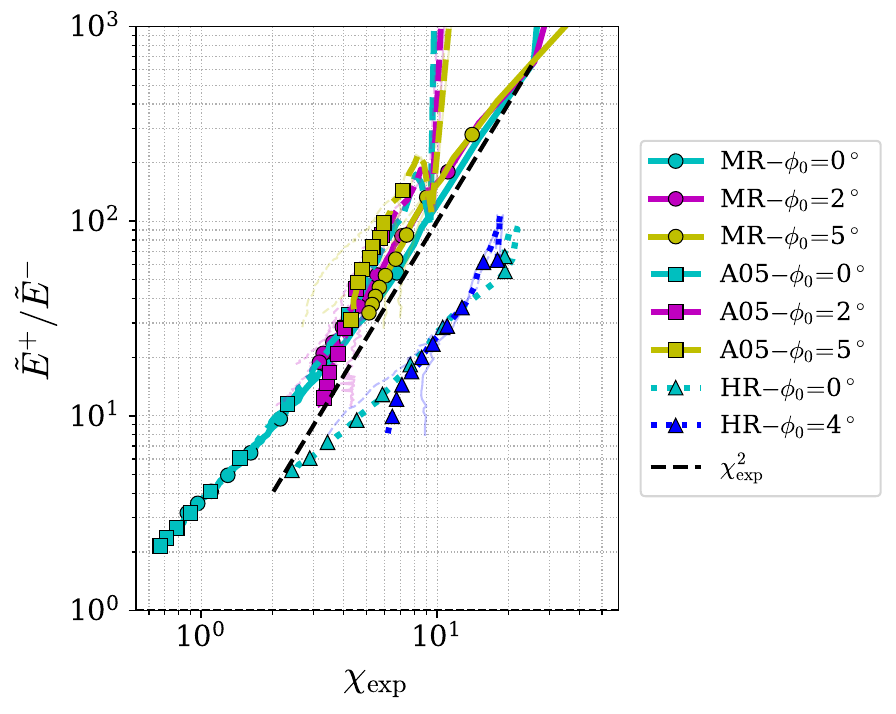}
    \caption{Ratio of wave energies $\tilde E^+/ \tilde E^-$ plotted against $\chi_{\rm exp}$ for all simulations (HR-$\Phi_0\!=\!0^\circ,4^\circ$ and MR-$\Phi_0\!=\!0^\circ,2^\circ,5^\circ$, A05-$\Phi_0\!=\!0^\circ,2^\circ,5^\circ$). The thick lines denote the perpendicularly averaged values for each case, while the faint solid and dashed curves correspond to the $\hat{\bm{e}}_T$ and $\hat{\bm{e}}_N$ components, respectively. The black dashed line indicates the theoretical scaling corresponding to the prediction from RDT theory (\cref{Phenomnology}). }
    \label{fig:energy ratio}
\end{figure}
One of the key predictions of the RDT model, which results directly from the balance between reflection and nonlinearity as opposed to depending on numerical coefficients, is that the energy ratio between counter-propagating Elsässer fields, $\tilde{E}^+/\tilde{E}^{-}$, should scale as $\chi_{\rm exp}^2$ (see Eq.~\ref{RDT Prediction}). This diagnostic should remove any differences between the PS and radial cases by folding the changes to $\ell_\perp(a)$ into $\chi_{\rm exp}(a)$; it thus acts as a direct test of the RDT phenomenology more generally.
Figure~\ref{fig:energy ratio} shows this ratio versus $\chi_{\rm exp}$ for all simulations. During the initial stage, the system undergoes a brief adjustment period. Afterward, the MR and A05 runs follow the nearly expected $\chi_{\rm exp}^2$ scaling trend. The HR runs exhibit systematically lower imbalance (smaller $(\tilde{z}^+)^2/(\tilde{z}^-)^2$), and all runs (including MR and A05) show a tendency to flatten at later times.  
The origin of this lower imbalance (and the late-time flattening) may be related to differences in the injected spectral parameters (e.g, $k_{\rm width}$ or $k_{\rm peak}$) or to resolution-dependent effects in the HR runs.
A more detailed investigation is required to isolate the physical mechanism underlying this behavior. Future work should include systematic resolution and domain-size scans, and controlled variations of the spectral parameters $k_w$ or $k_{\rm peak}$, although these cannot always be separated from $\chi_{\rm exp}$ and $A$ as independent parameters.

\subsection{Fluctuation Spectra}
\begin{figure}
    \centering
    \includegraphics[width=\linewidth]{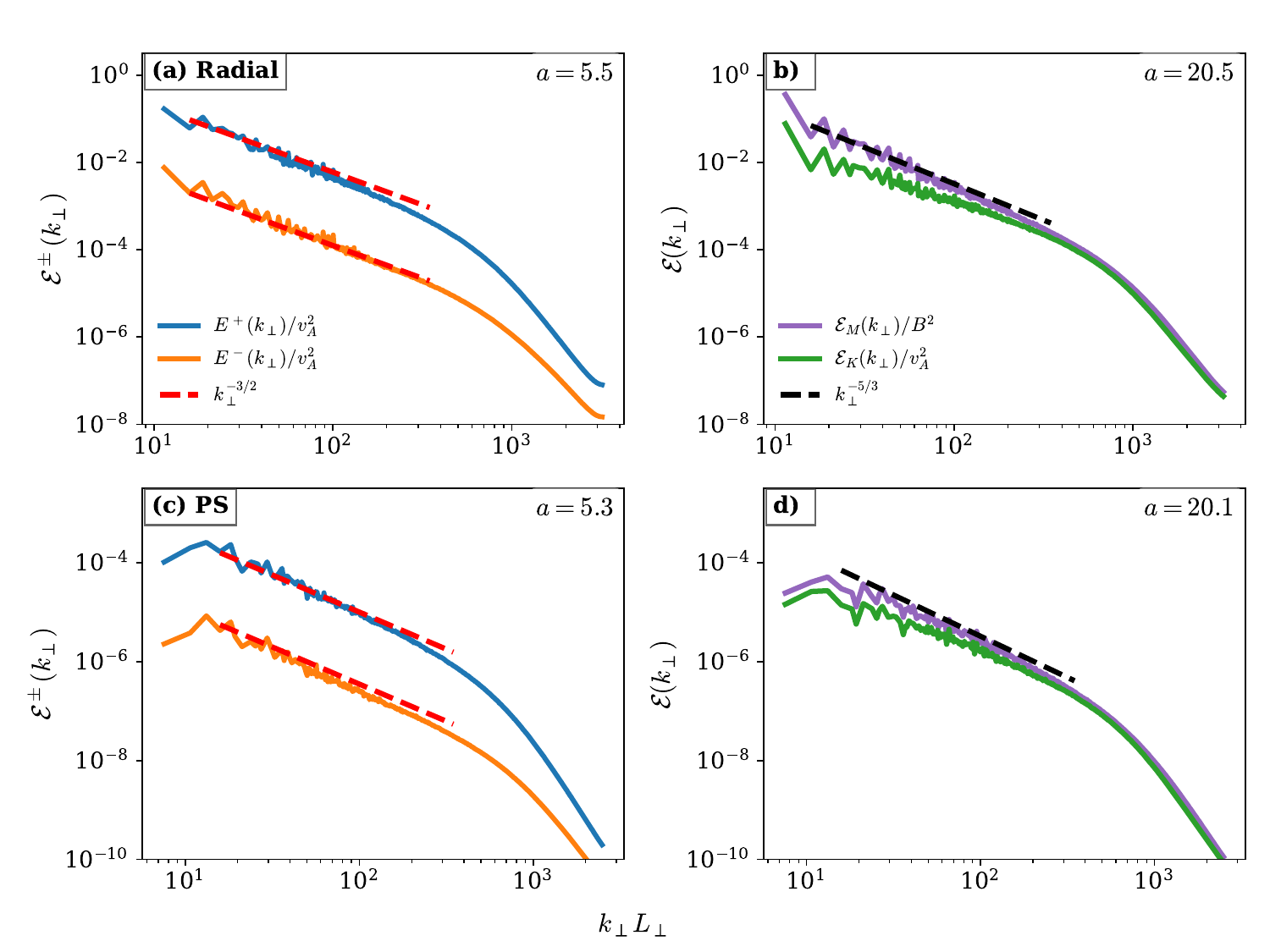}
\caption{Perpendicular energy spectra in the HR-$\Phi_0\!=\!0^\circ$ and HR-$\Phi_0\!=\!4^\circ$ simulations. All spectra are plotted versus dimensionless perpendicular wavenumber $k_\perp L_\perp$ ($L_\perp$ is the co-moving perpendicular box length).
Panel (a): Radial field evolution of outward and inward energy Elsässer spectra ${\mathcal{E}}^\pm$ at the indicated expansion times, showing the development of imbalance with scale. (b) At $a=20.5$ the magnetic ($\mathcal{E}_M$, purple) and kinetic ($\mathcal{E}_K$, green) spectra for the radial run; magnetic energy dominates across the inertial range. (c) HR-$\Phi_0\!=\!4^\circ$ Elsässer spectra at $a=5.5$, demonstrating a stronger imbalance. (d) Magnetic and Kinetic energy spectra for $\Phi_0\!=\!4^\circ$ at $a=20.1$, which display comparable inertial-range slopes.}
\label{fig:spectra}
\end{figure}
The perpendicular power spectra of outward and inward Elsässer fluctuations are shown in Fig.~\ref{fig:spectra} for the HR-$\Phi_0\!=\!0^\circ$ and HR-$\Phi_0\!=\!4^\circ$ runs. Single radial and PS snapshot at $a=5$ and $a\simeq 20$ is shown.
In the inertial range, the outward/inward cascade $\tilde{\mathcal{E}}^\pm$ follows a slope near $k_\perp^{-3/2}$. 
The Parker spiral snapshot shows nearly identical inertial-range shapes. Observational and numerical studies of perpendicular spectra and cross-helicity (e.g., \citealt{Podesta2009}) likewise find similar persistent slopes across a range of solar-wind conditions.
At late times, both simulations evolve toward magnetically dominated spectra, with the radial run showing stronger dominance, as expected based on the discussion above. 
  We also note that the dissipative range, where spectra steepen and eventually decay exponentially at the grid scale, appears similar in both geometries. This is expected, as the numerical dissipation mechanism in Athena++ operates identically regardless of the mean-field configuration. The geometric effects of the Parker spiral primarily influence the development and persistence of the outer scale of turbulence, rather than modifying the dissipation physics at the smallest resolved scales.  
Overall, the 1D perpendicular spectra appear largely insensitive to mean-field obliquity. This underscores that geometry primarily controls the extent and persistence of turbulence, rather than its local cascade law. 

These results also highlight the limitations of 1D spectral diagnostics. The apparent similarity of the $k_\perp$ spectra belies the more intricate, scale-dependent anisotropy evident in Fig.~\ref{fig:corr_lengths}, shaped by the spiral geometry. As observed by \cite{Verdini2018a}, global spectra retain an approximate $-5/3$ slope even when local, field-aligned structure functions reveal strong scale-dependent anisotropy. More detailed analyses---beyond 1D average spectra---could reveal whether the Parker spiral imprints distinctive anisotropy or phase structure below the outer scale. For now, the spectra simply confirm that both runs achieve strong Alfvénic turbulence with nearly Kolmogorov-like inertial ranges \citep{GS95}. 

\section{Other Turbulence Properties}\label{Observables}
Above, we have seen numerical support for theory outlined in \cref{Theory}. Here, we explore some more directly observable diagnostics, like switchback statistics and magnetic compressibility, which could provide potentially interesting ways to test our results, and RDT more generally, with in-situ spacecraft observations.

\subsection{Joint Evolution of Cross Helicity and Residual Energy}
\begin{figure}
    \centering
    \includegraphics[width=\textwidth]{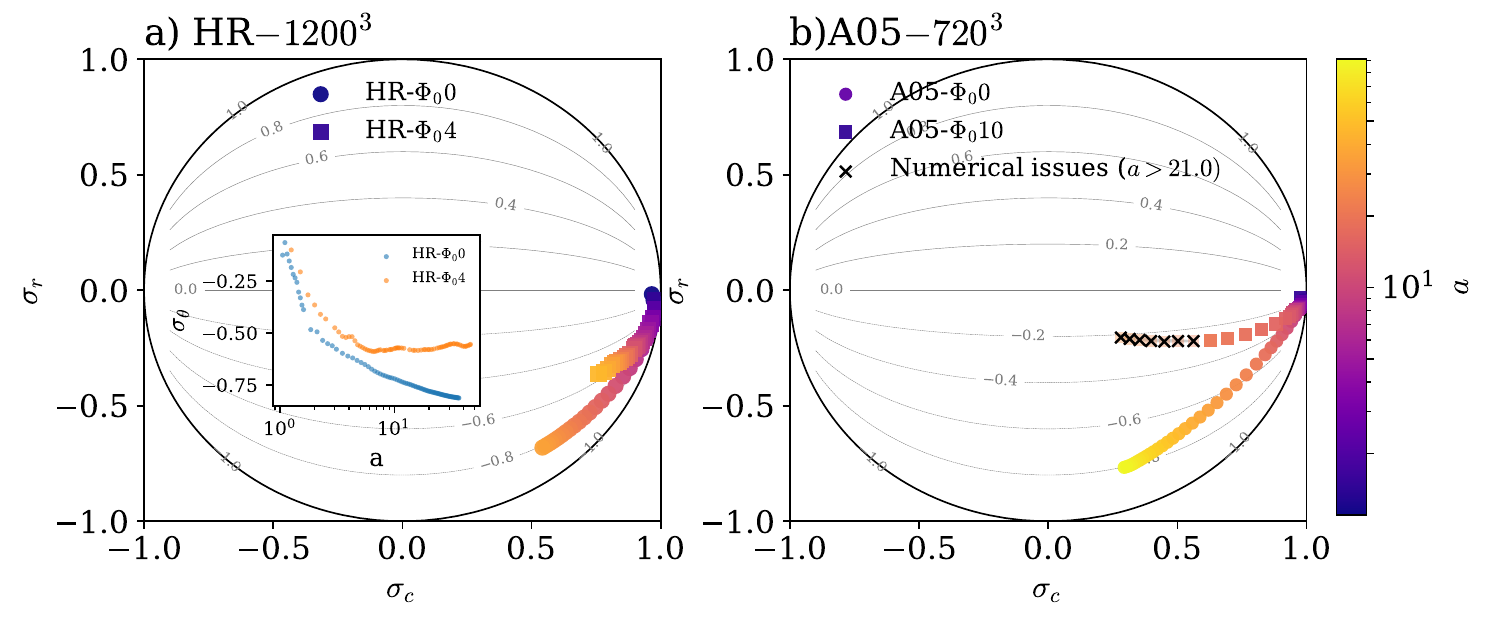}
    \caption{ Parametric evolution of cross helicity $\sigma_{c}$ and residual energy $\sigma_{r}$ as a function of expansion factor $a$. Panel (a) shows results from the HR-$\Phi_0\!=\!0^\circ,4^\circ$ simulations: circles correspond to the purely radial run and squares to the PS run. Panel (b) shows the A05-$\Phi_0\!=\!0^\circ,10^\circ$ simulations for both the radial and the higher initial Parker-angle case ($\Phi_0\!=\!10^\circ$).  Colored points represent $a = 1$ (dark blue), $a \sim 10$ (orange), and $a \sim 30$ (yellow). The black circle indicates the condition $\sigma_c^2 + \sigma_r^2 = 1$. Inset of (a) shows the alignment parameter $\sigma_\theta$ versus $a$ for HR–$\Phi_0=0^\circ$ (blue) and HR–$\Phi_0\!=\!4^\circ$ (orange) simulations. The radial run exhibits progressively stronger anti-alignment (more negative $\sigma_\theta$) with $a$, while the PS case flattens, indicating saturation of the alignment. In panel (b) for the PS case, points with $a>20$ are identified as affected by numerical instability: these points are plotted for completeness and the region they occupy is highlighted (highlighted, details in Appendix~\ref{appendixB}).}
    \label{fig:parametric_residual_energy}
\end{figure}

To quantify turbulence imbalance and the magnetic/kinetic partition, we track the normalized residual energy (Eq.~\ref{eq:sigma_r}) and cross helicity (Eq.~\ref{eq:sigma_c}). 
We also define the alignment parameter $\sigma_\theta$,
\begin{equation}
    \sigma_\theta = \frac{\langle \bm{z}^+ \cdot \bm{z}^- \rangle}
    {\sqrt{\langle |\bm{z}^+|^2 \rangle \, \langle |\bm{z}^-|^2 \rangle}}
    = \frac{\sigma_r}{\sqrt{1-\sigma_c^2}}\,,
\end{equation}
such that fully aligned fluctuations with $\sigma_\theta=1$ lie on the circle $\sigma^2_r + \sigma^2_c = 1$. Observations show fluctuations clustered near the lower-right quadrant of the unit circle, shifting from \(\sigma_c\!\sim\!1,\ \sigma_r\!\sim\!0\) close to the Sun to \(\sigma_c\!\sim\!0,\ \sigma_r\!\sim\!-1\) at larger radii \citep{Bruno2007,Bruno2013,Wicks2013}.
Figure~\ref{fig:parametric_residual_energy} illustrates the trajectory of $(\sigma_c,\sigma_r)$ as the solar wind expands (parametrized by the expansion factor $a$), for both the pure radial field case and the PS case. 

In both cases, the system begins highly imbalanced ($\sigma_c\approx1$, i.e.\ almost pure $z^+$) with $\sigma_r\approx0$. As $a$ increases, the curve stays close to the edge of the unit circle meaning $\sigma_r$ becomes increasingly negative. Physically, outward $\bm z^+$ fluctuations are reflected by the inhomogeneous wind into inward $\bm z^-$ that are nearly anti-aligned ($\bm{z}^-\approx -\bm{z}^+$), so this naturally generates turbulence with $\langle \bm z^-\cdot \bm z^+\rangle<0$ (hence $\sigma_r<0$, Eq.~\ref{eq:sigma_r}). This eventually produces magnetically dominated Alfvénic vortices (Fig.~\ref{fig:snaps720R}) as $\chi_{\rm exp}$ decreases and $\sigma_c$ begins to decline toward zero (balanced turbulence), with $\sigma_r$ near its maximal negative values. The results are consistent with RMHD simulations of \citet{Meyrand2025} and in situ wind observations, which is transitions from imbalanced to nearly balanced turbulence with magnetic dominance beyond 1 AU \citep{Bruno2007}.

The PS runs (squares) initially track the radial trend at small $a$ (the mean field is nearly radial there), but stay further from the circle edge at larger expansion. Even modest obliquity alters reflection and the polarization of the reflected component: assuming an Alfvénic $\bm z^+$ predominantly perpendicular to $\overline{\bm B}$, a reflected inward field $\bm z^-$ is no longer perfectly $-\bm z^+$ but instead a rotated combination of components due to the different sign of reflection in the $x$ direction (see Eq.~\eqref{eq:wave action}). This polarization mismatch reduces the correlation $\langle \bm z^+\!\cdot \bm z^-\rangle$, so $\sigma_r$ remains less negative at a given $\sigma_c$ (which also decays more slowly with $a$ for the PS case, as shown in the inset of Fig.~\ref{fig:energy_evolution}a).  
The inset of Fig.~\ref{fig:parametric_residual_energy}a shows $\sigma_\theta$ vs. $a$ for the HR-$\Phi_{0}\!=\!0^\circ$ and HR-$\Phi_{0}\!=\!4^\circ$ runs, quantifying the distance from the circle edge: HR-$\Phi_{0}\!=\!0^\circ$ trends toward substantially stronger anti-alignment with increasing $a$, whereas HR-$\Phi_{0}\!=\!4^\circ$ saturates at a smaller $|\sigma_\theta|$, which supports our view that the geometry-driven polarization mismatch between the $yz$ an $x$ directions reduces alignment.

These results suggest an observational test: the joint evolution of $\sigma_c$ and $\sigma_r$ could be examined as a function of local Parker spiral angle. The spiral angle, defined by Eq.~\eqref{eq:PS}, varies with distance above the ecliptic (heliocentric latitude) at fixed radius. So, one prediction is that the high-latitude turbulence should hug the edge of circle more closely than turbulence in the ecliptic.

\subsection{Switchbacks}\label{SBs}
\begin{figure}
    \centering
    \includegraphics[width=\linewidth]{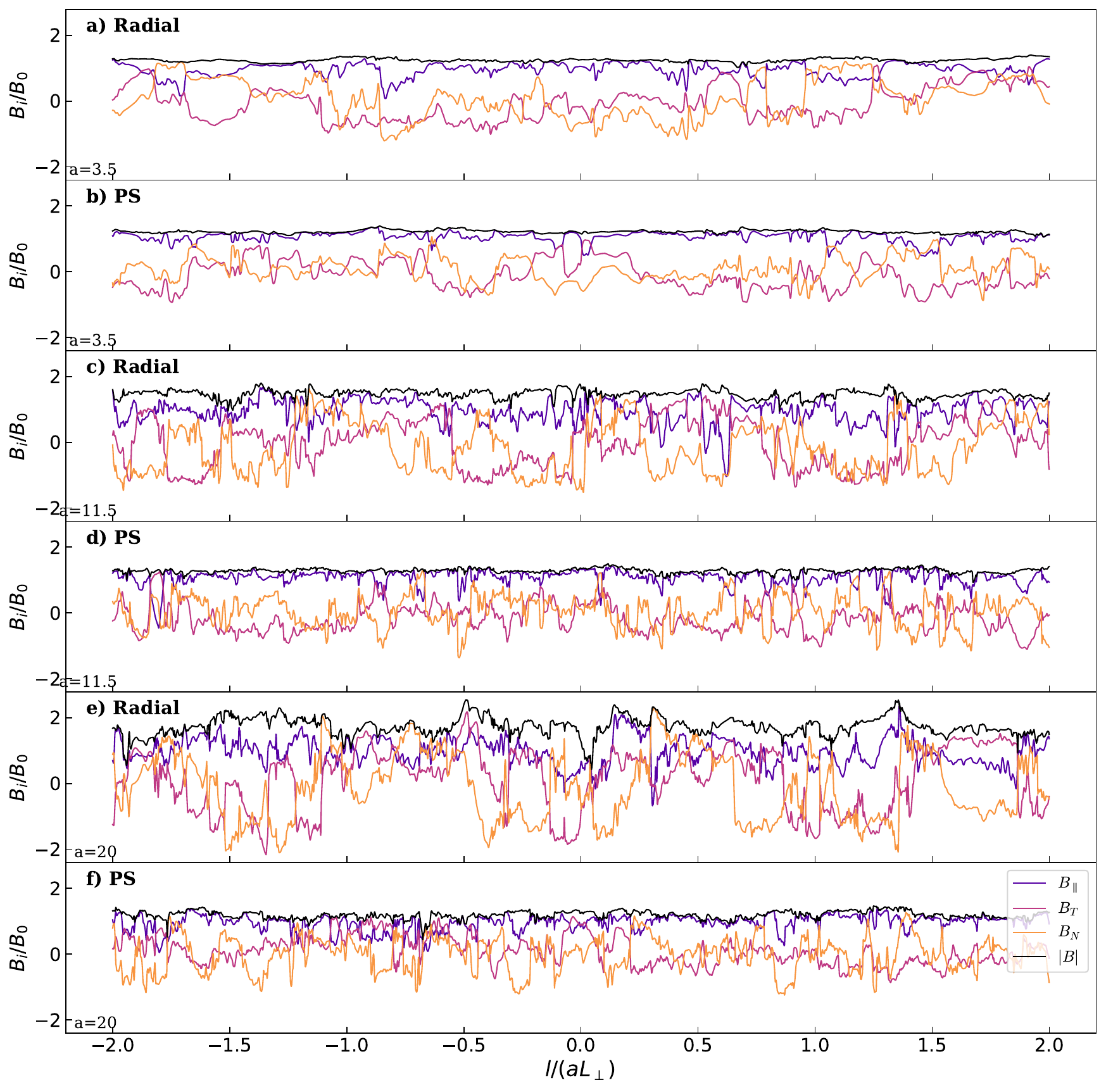}
    \caption{Fly-throughs of the magnetic field for the high-resolution radial (HR–$\Phi_0=0^\circ$ a,c,d) and PS (HR–$\Phi_0=4^\circ$, b,d,f) simulations at three expansion factors along the direction (1, 0.707, 0.392). Top row: a) radial and b) PS at $a=3.5$ ($\Phi \sim -13.75^\circ$); middle row: c) radial and d) PS at $a=11.5$ with $\Phi \sim 38.8^\circ$; bottom row: e) radial and f) PS at $a=20$ at $\Phi \sim -54.5^\circ$. In each panel the total field strength $|\bm{B}|$ is shown in black; $B_\parallel$, $B_T$, and $B_N$ are shown in blue, purple, and orange, respectively. Each component is plotted normalized to $\bar{B}$ (vertical axis $B_i/\bar B$) versus normalized distance $l/(aL_\perp)$, emphasizing how component behavior and $|\bm{B}|$ evolve with expansion.}
  \label{fig:flyby}
\end{figure}
Magnetic switchbacks (SBs) are sharp, transient reversals of the heliospheric magnetic field direction that occur with little change in field magnitude. First detected by Helios and later observed in abundance by PSP, they are now recognized as a ubiquitous feature of the near-Sun solar wind \citep{Bale2019,Kasper2019,Raouafi2023}. SBs are typically highly Alfvénic, their magnetic and velocity fields remain strongly correlated, and $|\bm B|$ is nearly constant throughout the reversal \citep{Bale2019}.

Figure~\ref{fig:flyby} illustrates representative fly-throughs at different stages of the evolution for both the radial and PS runs. In both geometries, $|\bm B|$ remains nearly constant while the field components undergo sharp directional swings in $B_\parallel$, accompanied by correlated deflections in $B_T$ and $B_N$. These signatures are consistent with large-amplitude, spherically polarized Alfvénic fluctuations as described in the past studies with similar setups \citep{Squire2020,Shoda2021,Johnston2022}. 
  The emergence of nearly constant-$|\bm B|$ fluctuations reflects the nonlinear evolution's tendency to favor spherically polarized, Alfvénic states that preserve field magnitude \citep{Barnes1974}. This is likely because any perturbation that changes $|\bm B|$ excites compressive (fast or slow magnetosonic) modes with associated density and pressure variations; these compressive components steepen into shocks and dissipate more rapidly than the incompressible Alfvénic part that maintains constant field magnitude. 
 
When $\delta B_\parallel$ reaches amplitudes comparable to the background, the local field direction can flip; the near constancy of $|\bm B|$ couples these radial reversals to transverse deflections, so each reversal appears as a coordinated multi-component swing. These rotations become noticeably sharper with expansion---compare the steeper swings between Fig.~\ref{fig:flyby}a and Fig.~\ref{fig:flyby}c---a natural consequence of the growing fluctuation amplitude. Furthermore, rotations in the PS runs are generally sharper than in the radial runs, as predicted by \citet{Squire2022}. 
At late times, the radial case exhibits the loss of coordinated, constant-\(|\bm{B}|\) rotations and increasingly irregular field-strength variations. These variations grow with distance, indicating that the magnetic field becomes less Alfvénic and more structurally complex as the cascade weakens. This late-time change broadly consistent with recent observational reports of enhanced field-strength variability with increasing heliocentric distance \citep{Horbury2023}. In contrast, the PS configuration retains a higher degree of Alfvénicity at comparable expansion factors, maintaining near-constant \(|\bm{B}|\) through large directional rotations and better preserving its nearly spherically polarized character.

\begin{figure}
    \centering
    \includegraphics[width=\linewidth]{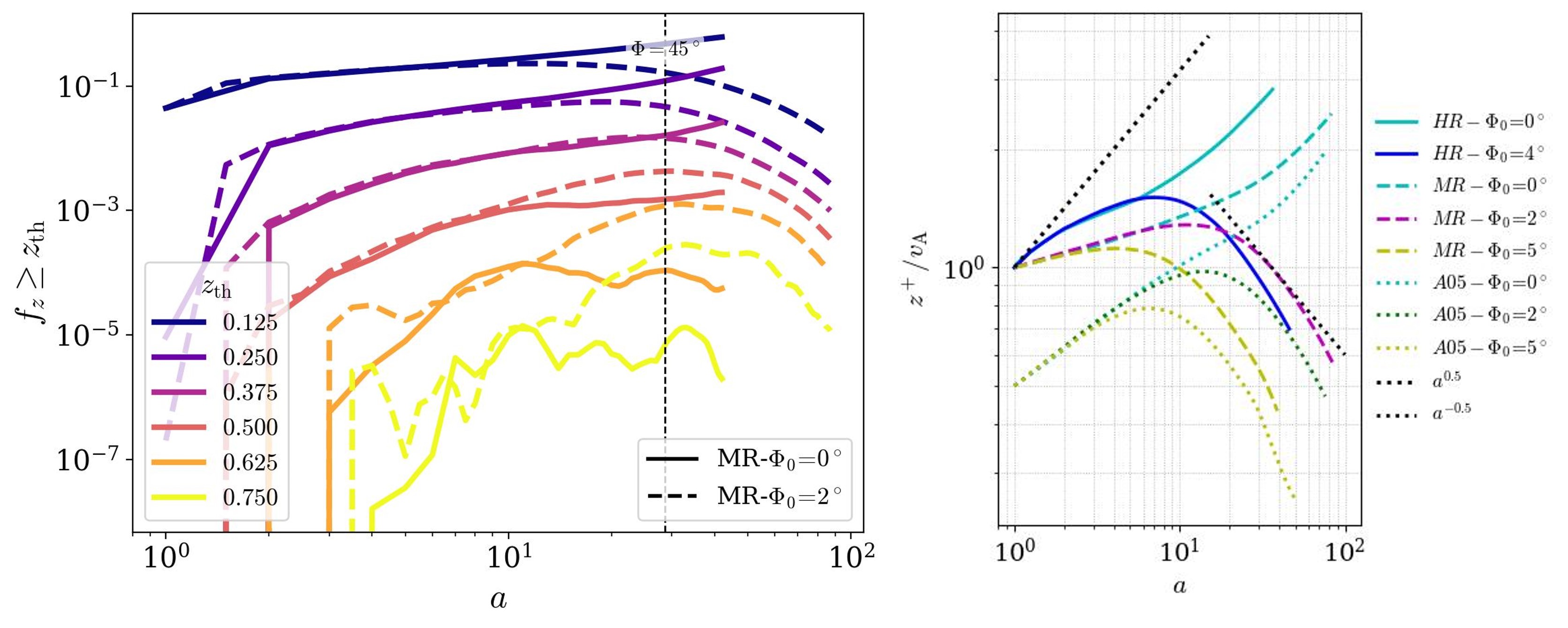}
    \caption{Evolution of switchback fraction ($f_{z} \ge z_{\rm th}$) in MR-$\Phi_0\!=\! 0^\circ$ (solid) and MR-$\Phi_0\!=\!2^\circ$ (dashed lines). Both runs produce switchbacks, but the PS case exhibits systematically larger switchback fractions across effectively all $a$ and a stronger growth of large-angle deflections with $a$ up to the point where fluctuation amplitudes decline; the downturn in ($f_{z} \ge z_{\rm th}$) at the largest $a$ is caused by the overall decrease in fluctuation amplitude. Right panel: the amplitude \(z^+/v_{\rm A}\) versus $a$ for all the simulations listed in the Table.~\ref{tab1:sim_params}. Dotted black lines show the WKB expectation for a radial field ($z^+/v_{\rm A} \propto a^{1/2}$) or azimuthally dominated field ($z^+/v_{\rm A} \propto a^{-1/2}$).}
  \label{fig:SB_Fraction}
\end{figure}

We examine the switchback fraction in our simulations, for which we define a switchback as a region where the magnetic field $\bm B$ deviates from the local mean field $\bar{\bm B} $ by more than a specified threshold angle. This deviation is quantified by the normalized deflection parameter 
\begin{equation}
    z = \frac{1}{2}(1-\cos{\vartheta_z}),
\end{equation}
where the deflection angle $\vartheta_z$ is given by
\begin{equation}
    \cos{\vartheta_z} = \frac{\bm B \cdot {\bar{\bm B}}}{\lvert\bm B\rvert \lvert{\bar{\bm B}}\rvert}
\end{equation}
Here, $z = 0$ corresponds to a field perfectly aligned with the background, and $z = 1$ to antiparallel configuration. We analyze regions corresponding to deflection thresholds $z_{\rm th} = 0.125,\;0.25,\;0.375,\;0.5,\;0.625,\;0.75$. We find that PS runs produce a larger fraction of strong directional reversals ($z\gtrsim0.5$), but a smaller fraction of weak deflections, than the corresponding radial runs (Fig.~\ref{fig:SB_Fraction}), in agreement with the 3D simulations of \citet{Johnston2022} and analytic explanation of \citet{Squire2022}. 
In the PS case, the switchback fraction grows until a near spiral angle of \(45^\circ\) and then declines. This decline coincides with the change in the evolution of \(v_{\rm A}\) between the two geometries discussed in \cref{Theory}. The right panel of Fig.~\ref{fig:SB_Fraction} shows the evolution of the normalized amplitude $z^+/v_{\rm A}$ for all the simulation sets which are compared to the WKB expectations for a radial ($z^+/v_{\rm A} \propto a^{1/2}$) or azimuthal ($z^+/v_{\rm A} \propto a^{-1/2}$) background field. In the radial case, the amplitude rise slowly with increasing $a$, whereas the PS runs show a different behavior: they remain approximately flat or grow slowly at small $a$ and then decline at large $a$. 
\subsection{Compressibility}
\begin{figure}
    \includegraphics[width=\linewidth]{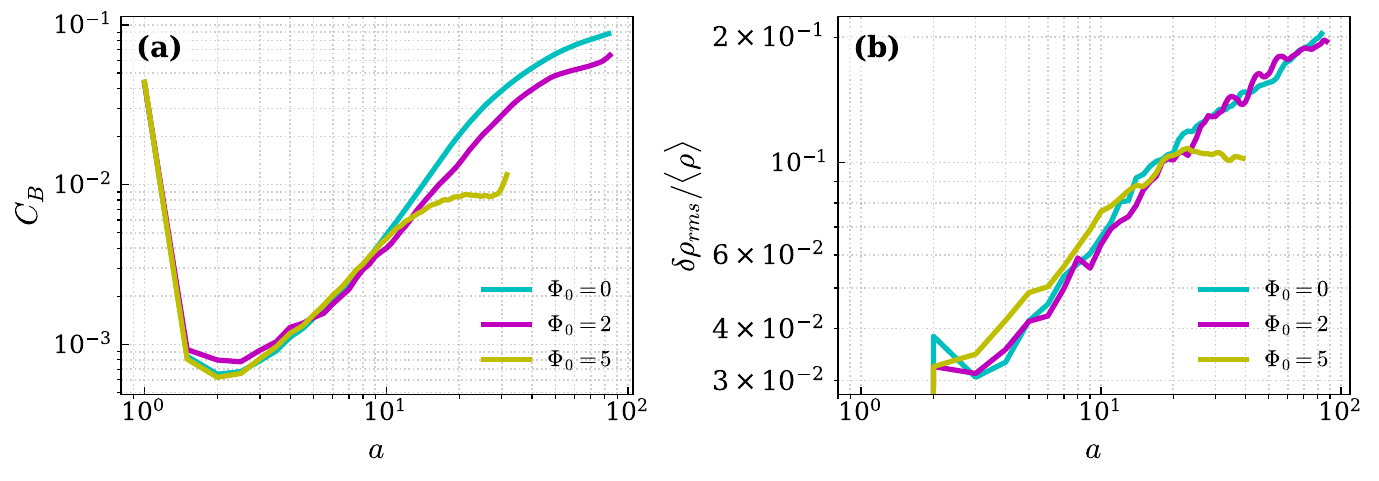}
    \caption{Evolution of (a) magnetic compressibility \(C_B\), and (b) density fluctuation amplitude \(\delta\rho_{\rm rms}/\langle\rho\rangle\) for the MR–\(\Phi_0\!=\!0^\circ,2^\circ,5^\circ\) simulations. All cases start with relatively large \(C_B\) because of the linearly polarized initial conditions.}
    \label{CMB}
\end{figure}

As noted in \Cref{SBs}, spherically polarized Alfvénic fluctuations are observed in the solar wind. To quantify the degree of compressibility in the evolving turbulence, we therefore measure the compressive fraction of the fluctuations.
We use the diagnostics
\begin{gather}\label{eq:mag_compress}  
{C_B = \sqrt{\frac{\langle(\delta|B|)^2\rangle}{\langle|\delta\bm{B}|^2\rangle}}} ,\quad  \frac{\delta\rho_\mathrm{rms}}{\langle\rho\rangle} = \sqrt{\left\langle \left( \frac{\rho - \langle \rho\rangle}{\langle \rho\rangle}\right)^2\right \rangle}.
\end{gather}
$C_B$ is nearly zero for pure transverse waves and grows when compressive fluctuations are present \citep{Chen2021, Shoda2021}. $\delta \rho_{\rm rms}/\langle\rho\rangle$ measures the fractional amplitude of density variations relative to the mean density. Taken together, $C_B$ and $\delta \rho_{\rm rms}$ measure the prevalence of compressive fluctuations in the system.
Figure~\ref{CMB} shows the evolution of $C_B$, and $\delta\rho_{\rm rms}/\langle\rho\rangle$ as functions of the expansion factor $a$. 
All runs start with higher $C_B$ because of the initialization of simulations with large-amplitude linearly polarized waves. As expansion and nonlinear mode coupling proceed, compressions in $|\bm B|$ rapidly dissipated within $\leq 1 \tau_{\rm A}$, driving the system toward spherical polarization and an effectively constant $|\bm B|$; correspondingly, $C_B $ drops significantly at early times.
 
These diagnostics show no clear evidence that the PS geometry directly drives stronger compressive activity. Rather, PS runs retain higher normalized cross-helicity, (Fig.~\ref{fig:energy_evolution}) producing sustained, near-constant-$|\bm B|$ rotations (Alfvénic switchbacks; Fig.~\ref{fig:flyby}). Compressive power increases as the turbulence becomes less Alfvénic---i.e. at lower $\sigma_c$ (or, equivalently, at larger $\sigma_r$)---as the relative contribution of $|\bm B|$-varying compressive fluctuations grows (see Fig.~\ref{fig:flyby}e). Therefore, as the system evolves from a strongly imbalanced state toward a more balanced regime it loses its spherical polarization and magnetic compressibility increases. Our simulations are consistent with this interpretation, with all cases similar at earlier times then differing later has $\sigma_c$ drops in the radial case, but remains higher for the PS ones (see Fig.~\ref{fig:energy_evolution}). 

We note an important caveat: the simulations presented here employ an   locally isothermal   equation of state, which does not capture the solar wind’s full thermodynamics. A more realistic equation of state should be included for direct quantitative comparison with spacecraft density signatures and compressive energetics.

\section{Conclusion}\label{conclusion}
We use high-resolution, three-dimensional expanding-box MHD simulations initialized with large-amplitude outward-propagating $\bm z^+$ fluctuations to explore how a non-radial (Parker-spiral) mean field alters reflection-driven turbulence (RDT) and its contribution to solar wind heating. Comparing radial and Parker spiral (PS) geometries, we test a simple RDT phenomenology, which is controlled by the ratio of expansion to nonlinear time, termed $\chi_{\rm exp}$, to explain the effect of PS and identify diagnostics that yield testable in-situ signatures. The radial and PS cases are broadly similar with RDT operating similarly in each case---the principal difference is the evolution of the perpendicular scale $\ell_\perp$, which has important consequences for the nonlinear time $\tau_{\rm nl}$, and therefore $\chi_{\rm exp}$. Our simulations thereby provide a useful, direct test of the RDT phenomenology and we see broad agreement between a simple phenomenology and the main trends in our simulations.

With a purely radial mean field the simulations show two phases. At small heliocentric distances the flow is strongly imbalanced \((z^+\gg z^-)\) and highly Alfvénic, while at larger radii the inward component grows, and the system approaches a more balanced state \((z^+\sim z^-)\). The dominant \(z^+\) component undergoes both forward (to smaller scales) and inverse (to larger scales) transfer, and the net result is a weakening of nonlinear interactions as \(\chi_{\rm exp}\) approaches 1. The phenomenology developed in \cref{Phenomnology}, which is effectively standard one of \citet{Dmitruk2002}, predicts \(z^+/z^-\sim\chi_{\rm exp}\); hence \(\chi_{\rm exp}\sim1\) marks the end of the imbalanced phase. During this late phase the Elsässer fields become strongly aligned $(\bm z^- \propto -\bm z^+)$, the flow organizes into quasi two-dimensional, magnetically dominated Alfvénic vortices, and volumetric dissipation is reduced. Our results are broadly consistent with the reduced MHD study of \citet{Meyrand2025}, extending them to a compressible framework. This also shows how constant-$|\bm B|$ spherically polarized switchbacks give way to more compressive balanced turbulence with larger $\lvert\bm B \rvert$ variation, a transition also observed in the solar wind \citep{Horbury2023}.

Introducing a Parker-spiral background produces longer-lived turbulent state and more heating, but otherwise similar properties. Geometrically, the azimuthal mean-field component breaks the cylindrical symmetry present in the radial case: as the plasma expands, eddies are stretched in the azimuthal and vertical directions, but this differs to the local field direction, producing anisotropic, ribbon-like structures with two distinct perpendicular correlation lengths \((\ell_{\perp,\rm T}\) tangential, \(\ell_{\perp,\rm N}\) normal). These transverse lengths are much smaller than the $\ell_\perp$ with a radial mean field because of the difference between field and expansion directions, and set the effective nonlinear outer scale. This causes the effective $\ell_\perp$ to initially grow with expansion, as with a radial field, but then saturate, meaning the nonlinear time no longer increases relative to the expansion time. This behavior causes $\chi_{\rm exp}$ to flatten at late times, extending the imbalanced phase of the nonlinear cascade and thereby continuing to dissipate energy at larger distances instead of freezing into weak vortical states as in the radial case (see §\ref{sub:PS Theory} and §\ref{theoretical validation}). 

We provide various additional diagnostics that could be interesting to connect to in-situ observations. The joint evolution on the circle plot $(\sigma_c,\sigma_r$; Fig.~\ref{fig:parametric_residual_energy})---from high $\sigma_c$, small $\sigma_r$ near the Sun toward reduced $\sigma_c$ and more negative $\sigma_r$ at larger radii---depends on the Parker spiral angle \citep{Bavassano1998,Bruno2007,Wicks2013,Meyrand2025}, with the PS making it stray further from the circles edge. Interestingly, the PS geometry does not systematically increase or decrease compressive activity; compressibility diagnostics (e.g., $C_B$, $\delta \rho_{\rm rms}/\langle\rho\rangle$) show no persistent enhancement in PS runs. Rather, the differences between PS and radial cases reflect that PS intervals remain more Alfvénic and, as a consequence, preserve near constant-$|\bm B|$ rotations to larger $a$. This yields generally sharper, more Alfvénic switchbacks for the PS geometry than in a purely radial field. Including a Parker spiral component tends to increase the growth and persistence of switchbacks with expansion, but the slower decay of $v_{\rm A}$ once $\Phi>45^\circ$ can halt that growth (Fig.~\ref{fig:SB_Fraction}). These predictions could be directly testable with spacecraft observations, perhaps via comparisons across heliocentric latitudes.

Our results are subject to several controlled approximations: they are obtained from compressible   locally isothermal   MHD simulations in an expanding-box framework with constant solar-wind speed. The model therefore applies most directly outside the Alfvén point for the parameter ranges explored here---processes that amplify fluctuations inside the Alfvén point (and the accelerating background) are not included and there is no PS in such a regime anyway. Accordingly, our simulations should be interpreted as the nonlinear evolution of large-amplitude outward $z^+$ perturbations that are already present at or beyond the Alfvén point. To study the early-stage amplification of waves, one must adopt frameworks that capture the accelerating wind \citep{Chandran2019,Ballegooijen2011,Tenerani2017,Shoda2021}. 

In addition to these caveats, we note several important limitations and directions for future work. First, the finite box size in the parallel direction ($L_{\parallel}$) constrains the longest-parallel wavelengths that can be represented and can therefore bias measured correlation lengths and decay rates; systematic domain-size scans are needed to quantify these effects. Second, the   locally isothermal   MHD approximation omits kinetic physics (e.g., collisionless damping, pressure anisotropy, and other wave-particle processes) that can modify both reflection and dissipation, so extensions to kinetic or hybrid models are desirable to assess how robust the present MHD results are. Third, some aspects of our conclusions depend on the assumed spectral parameters of the initial conditions (e.g. $k_{\rm width}$ and $\kappa$); the sensitivity of the dynamics to the initial spectral width should be explored with controlled variations. Finally, we have not fully explored the space of expansion-nonlinearity ratios: further runs with different $\chi_{\rm exp}$ and with higher initial fluctuation amplitudes are required to map the transition to stronger nonlinearity and to determine how the results scale with initial conditions.

\section*{Acknowledgements}
The authors gratefully acknowledge R. Meyrand, Z. Johnston, and B. D. G. Chandran for valuable discussions and insights that contributed to this work. This research was supported by the University of Otago through a University of Otago Doctoral Scholarship (K. A.), and by the Royal Society Te Apārangi through Marsden Fund grant MFP-UOO2221 (J. S.). Computational resources were provided by the New Zealand eScience Infrastructure (NeSI) under project grant uoo02637.

\section*{Declaration of interest}
The authors report no conflict of interest.

\appendix
\section{Linear Evolution}
To isolate the wave‐like dynamics that underlie our fully nonlinear simulations, we examine the \emph{linear} evolution of small Alfvénic perturbations in the expanding‐box framework. Assuming incompressibility, we linearize Eq.~\eqref{eq:final}, with $\nabla p$ chosen to enforce $\nabla\cdot\bm z^\pm =0$.
For a single Fourier mode we use the plane-wave ansatz;
\begin{equation}\label{eq:plane_wave_ansatz} 
\bm z_1^{\pm}(\bm r,t)=\bm z_1^{\pm}(t)e^{i\bm k(t)\cdot\bm r}, 
\end{equation} 
with expanding wavevector components, \begin{equation}\label{eq:k_of_a} \bm k(t) = (k_{x0},\;k_{y0}/a,\;k_{z0}/a) \equiv (k_x,\;k_y/a,\;k_z/a). \end{equation} Consequently, 
\begin{equation}\label{eq:dotk} 
\dot{\bm k}(t) = -\frac{\dot a}{a}\bm k_{\perp}(t),\qquad \bm k_{\perp}=(0,k_y,k_z). 
\end{equation}
The background Alfvén velocity is, \begin{equation}\label{eq:vA_of_a} \bm v_{\rm A}(t)=\Big(v_{{\rm A0}x}/a,\;v_{{\rm A0}y},\;0\Big), \end{equation} so the Alfvén frequency is \begin{equation}\label{eq:omegaA} \omega_{\rm A}(t)\equiv\bm v_{\rm A}(t)\cdot\bm k(t)=\frac{1}{a}(\bm k_0\cdot\bm v_{\rm A0}). \end{equation}
where $\bm k_0=(k_{x0},k_{y0},k_{z0})$ is the wave vector at $a=1$. Linearizing Eq.~\eqref{eq:final} and inserting the plane-wave ansatz yields (after dividing by the phase factor); \begin{equation}\label{eq:modal_start} \frac{d\bm z^{\pm}}{dt} \pm i(\bm v_A\cdot\bm k)\bm z^{\pm} + i\bm k p' = \bm S^{\pm}(t), \end{equation} where the expansion term is written compactly as, \begin{equation}\label{eq:Sdef} \bm S^{\pm}(t)= -\frac{\dot a}{2a}\mathbb T\cdot(\bm z^+ + \bm z^-)-\frac{\dot a}{2a}(z_x^{\pm}-z_x^{\mp}),\hat{\bm x}, \end{equation} with $\mathbb T=\mathrm{diag}(0,1,1)$.
Enforcing incompressibility mode‑by‑mode gives 
\begin{equation}\label{eq:incomp} 
    \bm k\cdot\bm z^{\pm}(t)=0, 
\end{equation} 
which yields 
\begin{equation}\label{eq:incomp_dt} 
\bm k\cdot\frac{d\bm z^{\pm}}{dt} = -\dot{\bm k}\cdot\bm z^{\pm}. 
\end{equation} 
Taking the dot product of (\ref{eq:modal_start}) with $\bm k$ and substituting (\ref{eq:incomp_dt}) gives the expression for the pressure scalar required to enforce incompressibility: 
\begin{equation}\label{eq:pressure} 
    p' = \frac{1}{i|\bm k|^2}\Big(\bm k\cdot\bm S^{\pm} + \dot{\bm k}\cdot\bm z^{\pm}\Big). 
\end{equation} 
Substituting (\ref{eq:pressure}) back into (\ref{eq:modal_start}) and rearranging yields the closed vector ODE \begin{equation}\label{eq:modal_nopressure} 
\frac{d\bm z^{\pm}}{dt} \pm i\omega_{\rm A}\bm z^{\pm} = \Pi \bm S^{\pm} - \bm k\frac{\dot{\bm k}\cdot\bm z^{\pm}}{|\bm k|^2}, \end{equation} 
where $\Pi=\mathbb I - \bm k\bm k/|\bm k|^2$ is the projection operator onto the plane transverse to $\bm k$. The last term on the right is parallel to $\bm k$ and therefore vanishes when the evolution is projected onto a transverse basis; nevertheless it contributes to the scalar pressure $p'$ via (\ref{eq:pressure}).
\begin{figure}
    \centering
    \includegraphics[width=\linewidth]{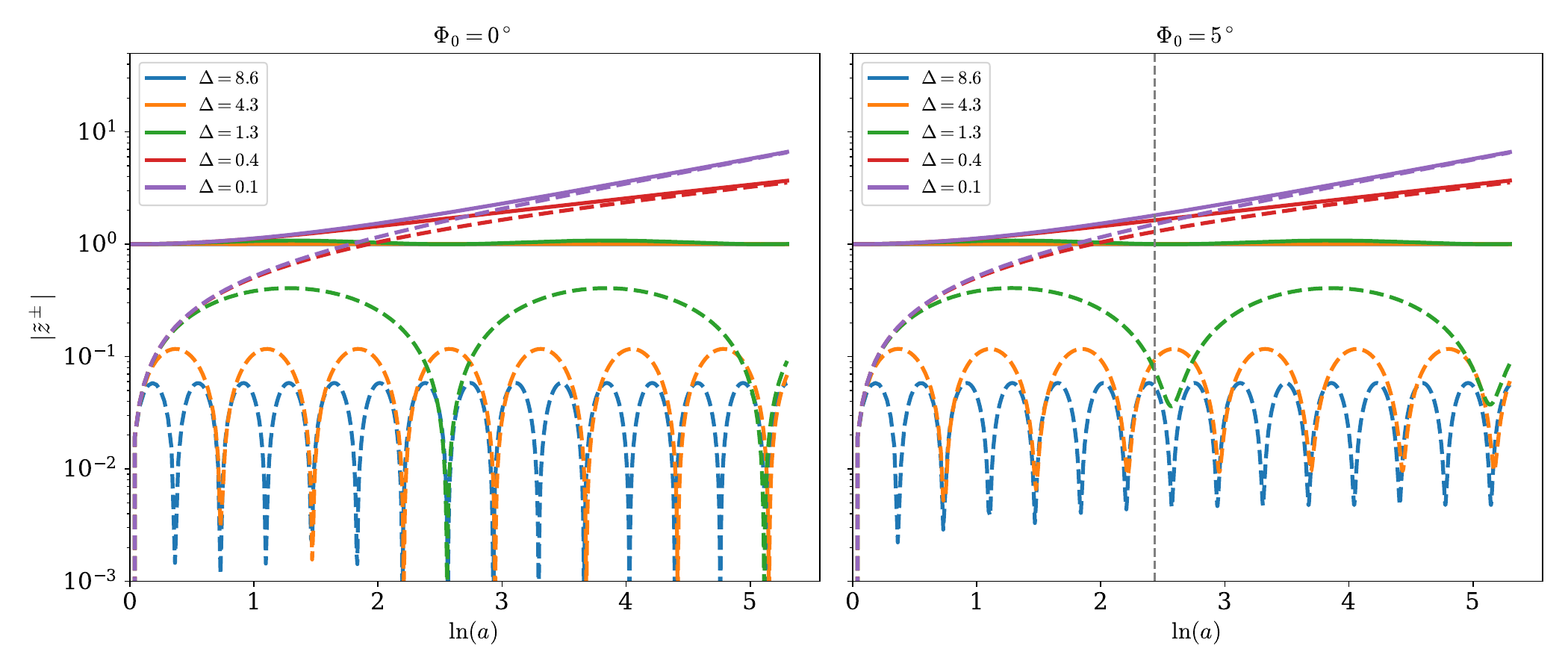}
    \caption{Solutions to the linear expanding incompressible MHD equations, for radial magnetic field ($\Phi_0 = 0^\circ$; left) and Parker spiral ($\Phi_0 = 5^\circ$; right). Initial conditions are a pure outward perturbation set by the normalized polarization $\tilde{\bm{z}}^+(a=1)=\bm k_0\times\bar{\bm B}$ and $\bm z^-(a=1)=0$. The other parameters used are ($|\bm k_0|=2\pi$, $\theta_{p0}=70^\circ$, $\varphi =90^\circ$), but results don't depend strongly on these choices. Solid lines show the outgoing component $|\bm z^+|$ and dashed lines the inward component $|\bm z^-|$; colors label different $\Delta \in\{8.6,4.3,1.3,0.4,0.1\}$ (legend). The grey vertical dashed line marks $\Phi\!=\!45^\circ$ as a guide. }
    \label{Linear_res}
\end{figure}

It is convenient to use $s=\ln a$ as the independent variable. Since $d/dt=(\dot a/a)\partial_{\ln a}$, multiplying (\ref{eq:modal_nopressure}) by $a/\dot a$ and simplifying yields the component‑wise evolution in the $x,y,z$ basis. Writing vectors in terms of the initial wavevector $\bm k_0$ and using \eqref{eq:dotk} one obtains,
\begin{multline}\label{eq:log_evol_components}
\partial_{\ln a}\begin{pmatrix} z_x^{\pm}\\ z_y^{\pm}\\ z_z^{\pm} \end{pmatrix}
= \mp \frac{i}{\dot a}(\bm k_0\cdot\bm v_{A0})\begin{pmatrix} z_x^{\pm}\\ z_y^{\pm}\\ z_z^{\pm} \end{pmatrix}
- \tfrac12\begin{pmatrix} z_x^{\pm} - z_x^{\mp}\\ z_y^{\pm} + z_y^{\mp}\\ z_z^{\pm} + z_z^{\mp} \end{pmatrix}\\
+ \frac{\bm k(t)}{2|\bm k(t)|^2}\Big[ 2\,\bm k_{\perp}(t)\cdot\bm z^{\pm} + k_{x0}(z_x^{\pm} - z_x^{\mp}) + \bm k_{\perp}(t)\cdot(\bm z^+ + \bm z^- )\Big].
\end{multline}
The $+2\,\bm k_{\perp}\cdot\bm z^{\pm}$ term arises from combining the $\dot{\bm k}\cdot\bm z^{\pm}$ term in the pressure numerator with the $a/\dot a$; the remaining terms follow straightforwardly from $\Pi\bm S^{\pm}$. 

As discussed in the main text, the key parameter governing the nature of linear fluctuations is the ratio,
\begin{equation}
\Delta \equiv \frac{\bm{k}(a) \cdot \bm{v}_A(a)}{\dot{a}/a}
= \frac{k_{x0} v_{A0x} + k_{y0} v_{A0y}}{\dot{a}}.
\end{equation}
Owing to the mixed scaling of the PS components, this is independent of $a$. We allow the wavevector $\bm k$ to have components in all three directions in order to sample the full range of geometrical orientations. We therefore parameterize
\begin{equation}
    \bm{k}(a)=k_0\bigl(\cos\theta_{p0},\; a^{-1}\sin\theta_{p0}\cos \varphi,\; a^{-1}\sin\theta_{p0}\sin \varphi\bigr),
\end{equation}
With this convention $(\varphi=\pm\pi/2)$ places $\bm{k}$ in the plane perpendicular to the PS mean field, whereas $(\varphi=0)$ or $(\varphi=\pi)$ places $\bm{k}$ in the same plane as the Parker spiral (\cref{Phenomnology}). \\
Figure~\ref{Linear_res} illustrates solutions to \eqref{eq:log_evol_components}  behavior for $\Phi_0=0$ and $\Phi_0=5^\circ$. Initial conditions are a pure outward wave with $\bm z^+(a=1)=(\bm k_0 \times \bar{\bm B})$ normalized to unit amplitude and $\bm z^-(a=1)=0$. In the PS solutions, the only model
change is the mean-field tilt $\Phi_0$, which introduces a nonzero transverse Alfvén component $v_{\rm{A}0\textit{y}}=v_0\sin{\Phi_0}$. This modifies the pressure/projection term, and the relevance of the modified reflection term in the $x$ direction, since $z^+_x=0$ for a radial field.

For \(|\Delta|>0.5\) modes, the Alfvén propagation term dominates and the outward fluctuation \(\tilde z^+\) propagates with an oscillatory phase and almost constant amplitude while the reflected component \(\tilde z^-\) remains small and oscillatory: \(|\tilde z^-|\) alternates in sign and does not grow secularly because the restoring Alfvén force counters the expansion term. By contrast, for long-wavelength (expansion-dominated) modes with \(|\Delta|<0.5\), the expansion terms overwhelm the Alfvénic restoring force, the mode becomes non-oscillatory (overdamped), and outward energy is steadily converted into inward fluctuations, with $z^-$ attaining a finite or slowly growing amplitude comparable to $z^+$ \citep{Meyrand2025}. 

Interestingly, the results obtained for the Parker spiral configuration are almost identical to those in the purely radial case. As a result, the linear dynamics in the Parker spiral do not introduce qualitatively new features beyond those already present in the radial-field case. This could partially explain why RDT behaves similarly in both cases, other than the evolution of $\ell_\perp$.

\section{Numerical artifacts}\label{appendixB}
\begin{figure}
        \centering
        \includegraphics[width=0.425\linewidth]{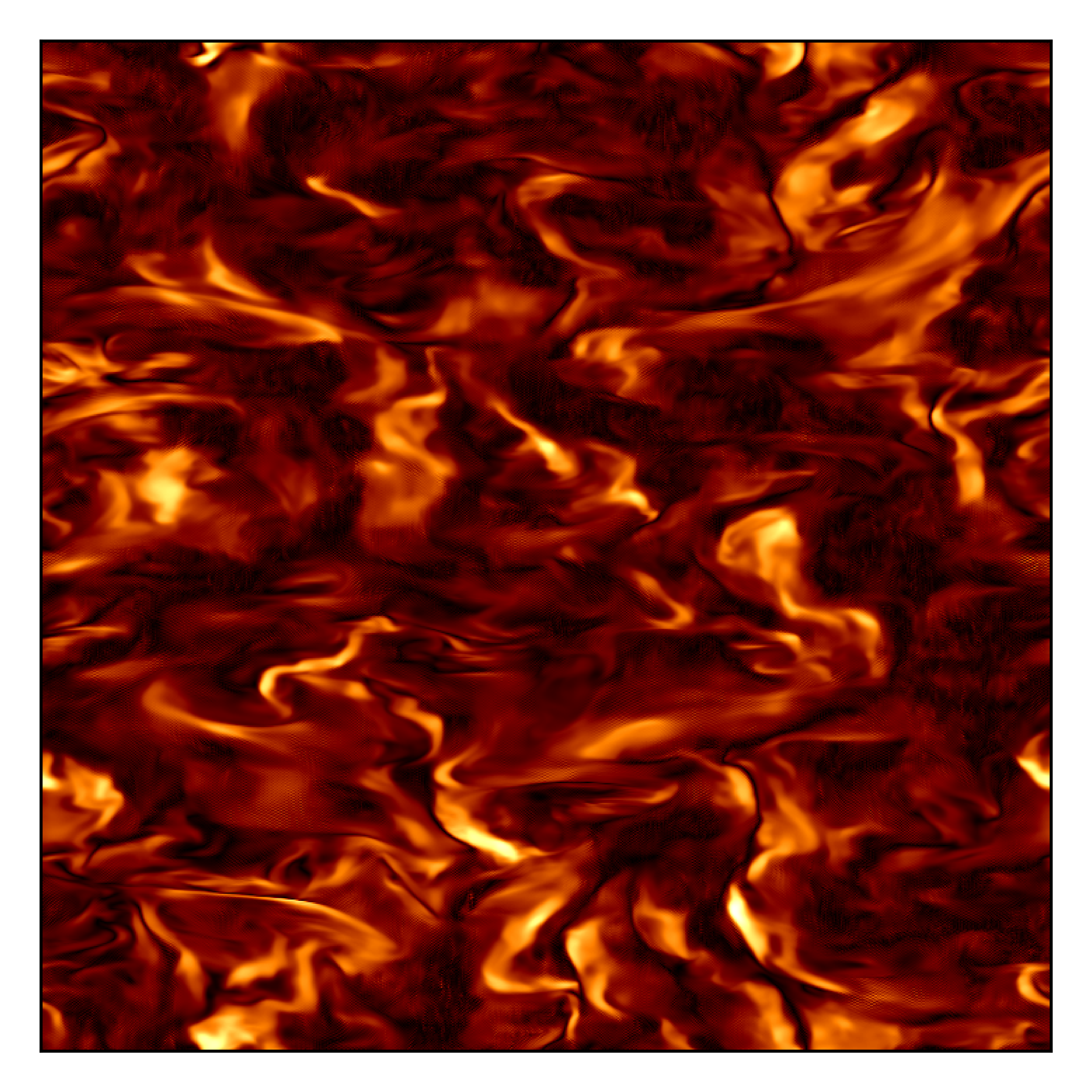}
        \includegraphics[width=0.425\linewidth]{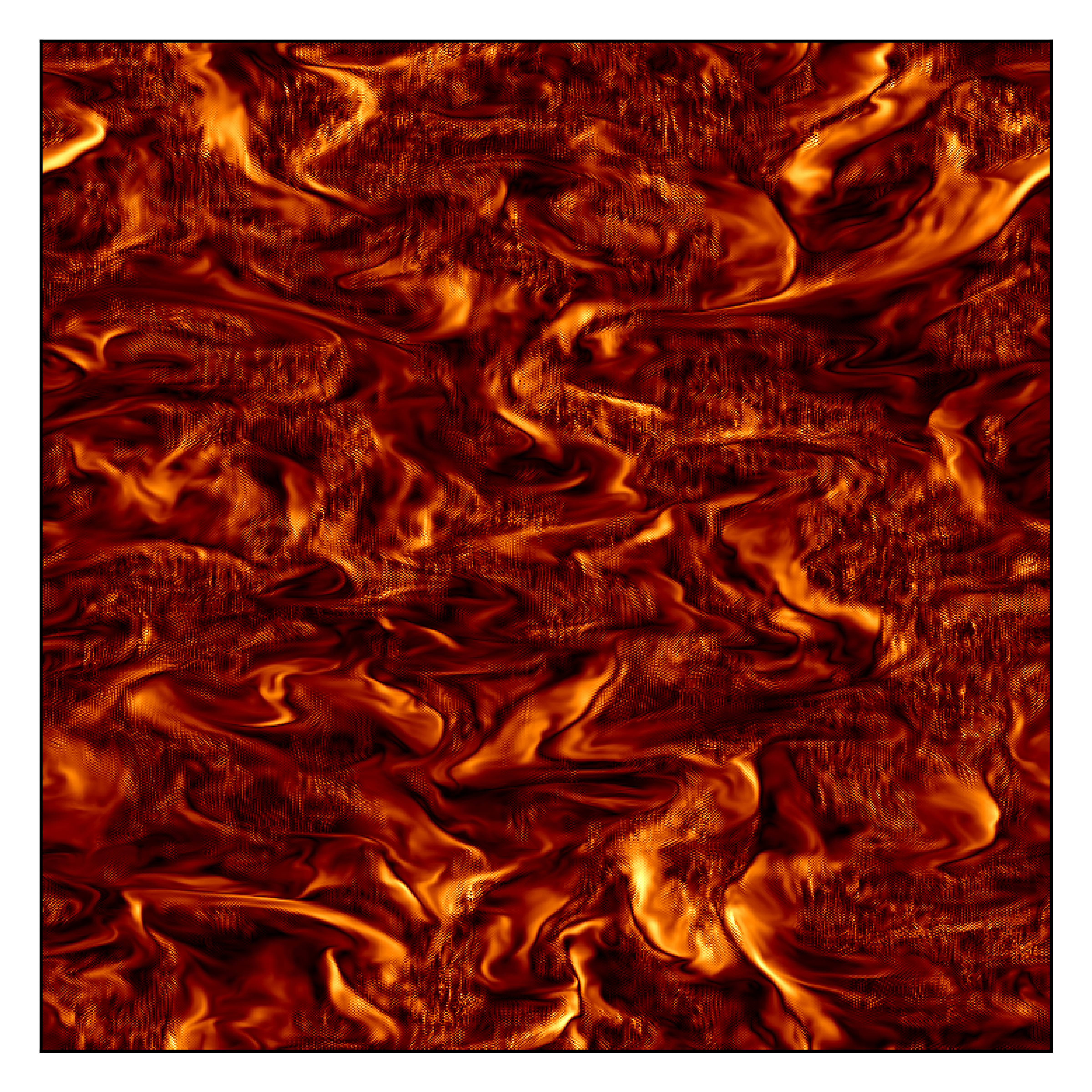}
\caption{Snapshot of $|\bm{z}^\pm_\perp|/|\bm{z}^\pm_\perp|_{\rm rms}$ in the $y$–$z$ plane at $a=39$ for MR-$\Phi_0\!=\!5^\circ$ run showing the onset of grid-scale, speckle-like noise at large spiral angle.}
\label{fig:appendix2}
\end{figure}

\begin{figure}
    \centering
    \includegraphics[width=\linewidth]{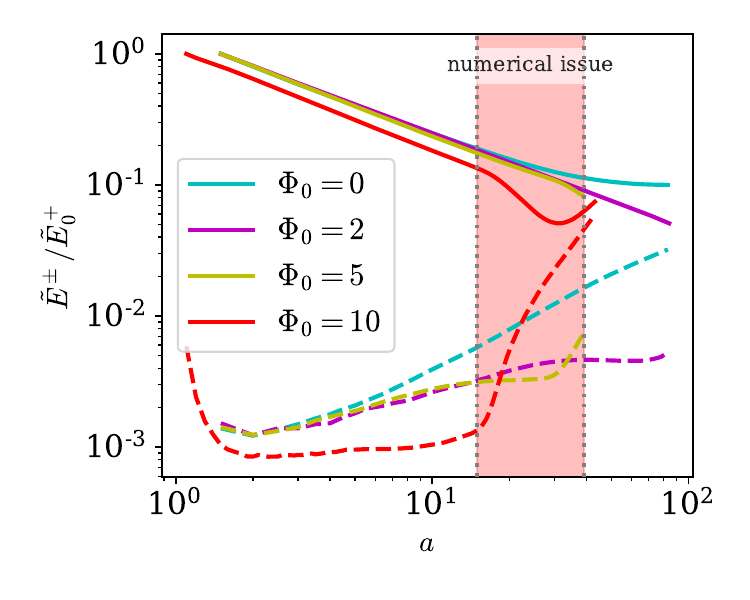}
    \caption{Time series of the outward energy for the MR-$\Phi_0\!=\!0^\circ,2^\circ,5^\circ$ runs. The red curve is an additional run with $k_{\rm peak} = (3.0, 3.0)$ and ($\Phi_0 =10^\circ$), included to illustrate the behavior at larger expansion $a$ ($\Phi \gtrsim 70^\circ$). The red shaded  region marks the time at which inward energy increases abruptly for $\Phi_0\!=\!10^\circ$ run.}
    \label{fig:appendix3}
\end{figure}

In the Parker spiral runs, a numerical instability usually emerges once the mean-field spiral angle exceeds $\sim70^\circ$. The onset is marked by the appearance of grid-scale, speckle-like noise in transverse Elsässer fields and by an abrupt, apparently nonphysical spike in the inward Elsässer energy $\tilde{E}^-$. A representative snapshot and the corresponding time series are shown in Fig.~\ref{fig:appendix2} and Fig.~\ref{fig:appendix3}, respectively.

A couple of lines of evidence suggest that this behavior is numerical rather than physical: (i) the energy jump is abrupt and coincides with the visual appearance of grid-scale noise; (ii) its reproducible occurrence at a similar spiral angle across multiple runs and resolutions, combined with the lack of any similar transition in the linear case (App. A). The instability is apparently absent in the purely radial configuration for the same runtime and parameter set, which argues against a simple expansion-only origin. The results could be related to the inability of the highly anisotropic computational domain to resolve small perpendicular structures.
In particular, the PS geometry at large spiral angle strongly anisotropises the flow, so that effective perpendicular gradients steepen. This reduces the effective resolution of transverse structures, perhaps causing the numerical issues.

Given the persistent distortion of large-scale statistics beyond the empirical threshold, we exclude data for which the mean-field spiral angle exceeds the reported threshold (approximately $70^\circ$) from the quantitative analysis presented in the main text. The exclusion boundary is set by inspecting snapshots, energy growth, and spectra. We emphasize that the threshold is empirical and that the potentially interesting physics that occurs at larger angles would be an important subject for future work.

\bibliographystyle{jpp}
\bibliography{references}
\end{document}